\documentclass[lettersize,journal]{IEEEtran}
\IEEEoverridecommandlockouts
\usepackage{cite}
\usepackage{verbatim}
\usepackage{amsmath,amssymb,amsfonts,bm}
\usepackage{algorithm}
\usepackage{algorithmicx}

\usepackage{bbm}
\usepackage{enumitem}
\usepackage{amsthm}
\usepackage{algpseudocode}
\usepackage{textcomp}

\usepackage{float}
\usepackage[font=small]{caption}
\usepackage{xcolor}
\usepackage{wrapfig}
\usepackage{hyperref}
\usepackage[capitalize]{cleveref}
\usepackage[font=footnotesize,labelformat=simple]{subcaption}
\usepackage[protrusion=true,expansion=true]{microtype}
\pdfoutput=1

\usepackage{caption}
\usepackage{multirow}
\usepackage{graphicx}
\everymath{\displaystyle}

\newtheorem{example}{Example}
\newtheorem{remark}{Remark}

\newcommand{\calC}{\mathcal{C}}

\newcommand{\calD}{\mathcal{D}}

\newcommand{\calG}{\mathcal{G}}

\newcommand{\calN}{\mathcal{N}}

\newcommand{\calZ}{\mathcal{Z}}
\newcommand{\bfU}{\mathbf{U}}

\newcommand{\bfb}{\mathbf{b}}

\newcommand{\bfD}{\mathbf{D}}

\newcommand{\bfF}{\mathbf{F}}

\newcommand{\bfV}{\mathbf{V}}

\newcommand{\bfW}{\mathbf{W}}
\newcommand{\bfA}{\mathbf{A}}

\newcommand{\bfT}{\mathbf{T}}

\newcommand{\bfs}{\mathbf{s}}

\newcommand{\bfM}{\mathbf{M}}
\newcommand{\bfR}{\mathbf{R}}

\newcommand{\bfP}{\mathbf{P}}
\newcommand{\bfQ}{\mathbf{Q}}

\DeclareMathOperator{\Tr}{Tr}

\hypersetup{draft}

\def\BibTeX{{\rm B\kern-.05em{\sc i\kern-.025em b}\kern-.08emT\kern-.1667em\lower.7ex\hbox{E}\kern-.125emX}}
    
\makeatother
\begin{document}


\title{Multi-Agent Reinforcement Learning for Graph Discovery in D2D-Enabled Federated Learning}

\author{
\IEEEauthorblockN{Satyavrat Wagle$^1$, Anindya Bijoy Das$^2$, David J. Love$^1$, and Christopher G. Brinton$^1$}

\IEEEauthorblockA{$^1$Elmore Family School of Electrical and Computer Engineering, Purdue University, $^2$University of Akron}
}

\maketitle

\begin{abstract}
Augmenting federated learning (FL) with device-to-device (D2D) communications can help improve convergence speed and reduce model bias through local information exchange. However, data privacy concerns, trust constraints between devices, and unreliable wireless channels each pose challenges in finding an effective yet resource efficient D2D graph structure. In this paper, we develop a decentralized reinforcement learning (RL) method for D2D graph discovery that promotes communication of impactful datapoints over reliable links for multiple learning paradigms, while following both data and device-specific trust constraints. An independent RL agent at each device trains a policy to predict the impact of incoming links in a decentralized manner without exposure of local data or significant communication overhead. For supervised settings, the D2D graph aims to improve device-specific label diversity without compromising system-level performance. For semi-supervised settings, we enable this by incorporating distributed label propagation. For unsupervised settings, we develop a variation-based diversity metric which estimates data diversity in terms of occupied latent space. Numerical experiments on five widely used datasets confirm that the data diversity improvements induced by our method increase convergence speed by up to $3 \times$ while reducing energy consumption by up to $5 \times$. They also show that our method is resilient to stragglers and changes in the aggregation interval. Finally, we show that our method offers scalability benefits for larger system sizes without increases in relative overhead, and adaptability to various downstream FL architectures and to dynamic wireless environments.\let\thefootnote\relax\footnote{This work was supported in part by the National Science Foundation (NSF) under grants CNS-2212565 and CPS-2313109, the Defense Advanced Research Projects Agency (DARPA) under grant D22AP00168, and the Office of Naval Research (ONR) under grant N00014-21-1-2472. 

This paper has supplementary downloadable material available at http://ieeexplore.ieee.org., provided by the author. The material includes Appendices A-D. Contact wagles@purdue.edu for further questions about this work.}
\addtocounter{footnote}{-1}\let\thefootnote\svthefootnote
\end{abstract}

\vspace{-0.05in}
\section{Introduction}\label{intro}

Federated learning (FL) has become a popular approach for global machine learning (ML) model construction across a set of distributed edge devices. The standard operation of FL consists of a coordinating server periodically aggregating models trained locally at edge devices on their respective local datasets. One of the fundamental challenges in FL is the presence of non-i.i.d data distributions across participating devices, which can slow convergence speed and result in global model bias~\cite{fed_to_fog}. These issues are exacerbated when some devices can only communicate their model updates to the server intermittently, e.g., due to poor channel conditions.

Recent studies suggest that device-to-device (D2D) communications that enable inter-device offloading of data processing is fast becoming the norm for distributed learning systems \cite{6g_survey,d2d_yess}. A recent trend of work has considered mitigating the challenges faced by FL systems by augmenting them with D2D communications in relevant network settings, e.g., wireless sensor networks~\cite{drl_multiuser_wsn}. In D2D-enabled FL, short-range information exchange is employed to reduce the tendency of devices to overfit on their locally collected datasets~\cite{wang2024device}. However, there are two factors which have a strong impact on the efficacy of such procedures: (i) \textit{inter-device trust and privacy concerns}, which may prevent data sharing between specific device pairs, possibly restricted to certain data classes; (ii) \textit{D2D wireless condition variations}, which impact communication efficiency and can result in intermittent data transmission failures.




A few studies have explored bias reduction in FL models through D2D information exchange. For example, they have considered offloading of (i) partial data sets to compensate for heterogeneous computation capabilities across devices \cite{fed_to_fog}, (ii) data to devices which are estimated to contribute more to system performance \cite{wang2024device}, and (iii) unlabeled data for decentralized pseudo-labeling \cite{iotmal}. The aforementioned works, however, do not consider the impact of communication reliability and inter-device trust in their implementations. The methods in \cite{flrl_dev_sel,flrl_dev_sel2} utilize reinforcement learning (RL)~\cite{sutton2018}  for training policies at the server to select devices for aggregation that reduce the bias of the system model. To the best of our knowledge, an RL based methodology to allow for \textit{device-level decision-making} to facilitate device-level cooperation in the presence of trust constraints and variable communication channels has not been studied. In addition, all of the above studies assume a centralized decision-making system, which exposes additional device information to the network.  

In order to address this gap, in this paper, we introduce an inter-device cooperation framework for D2D-enabled FL systems. Our method discovers critical inter-device links in D2D enabled federated learning over which convergence-critical information can be shared between devices while abiding by predefined inter-device trust constraints. We design a decentralized RL algorithm which allows for each device to create links independently while being cognizant of overall system performance as well as communication reliability. Our method is compatible with supervised, semi-supervised and unsupervised learning paradigms, and works in tandem with popular federated averaging algorithms. By employing decentralized decision-making in a system with centralized model aggregation, devices can share information between themselves without exposing any data-related information to the server. As the decision-making system fully bypasses the server, though, we can also naturally extend our method to utilize centralized decision-making in systems where data exchange with servers is allowed, or adapt it to decentralized model aggregations using consensus mechanisms.



\textbf{Outline:} The remainder of this paper is organized as follows. Section \ref{sec:system_model} compares and differentiates our work from existing literature, followed by a summary of contributions. In Section \ref{sec:problem_formulation}, we set up the system model for our framework and formalize the graph discovery problem as a diversity gain maximization over a data partitioning procedure and data diversity metric. Next, in Section \ref{sec:graph_formation}, we discuss the proposed decentralized RL method to solve the graph discovery problem in detail, cognizant of inter-device trust, communication reliability and overall system performance. In Section \ref{sec:idc}, we describe the data partitioning, message passing and data diversity calculations for supervised, semi-supervised and unsupervised settings. Finally, in Section \ref{sec:results}, we evaluate the performance of our method against baselines on five datasets, showing significant improvements in model training quality and energy consumption required to achieve performance benchmarks. 




\begin{figure}[t!]
    \centering
    \includegraphics[width=0.8\columnwidth]{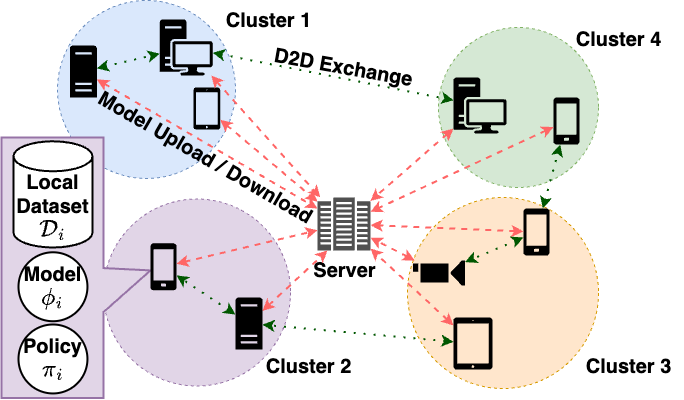}
    \caption{System model for D2D-enabled FL. In our RL-based approach for graph discovery, each device acts as a learning agent. Devices are partitioned into clusters of reliable devices to minimize probability of failure in communication.}
    \label{fig:sys_arch}\vspace{-3mm}
\end{figure}        

\section{Related Work and Summary of Contributions}
\label{sec:system_model}


\subsection{Related Work}



\subsubsection{Improving Convergence in D2D-Enabled FL}
D2D-enabled federated learning leverages inter-device communication in a federated system to improve convergence by  designing a network-aware variant of decentralized gradient descent \cite{d2d_comm_2}, mitigating model divergence by D2D exchange of model parameters aided by consensus mechanisms \cite{semidecentralized}. In \cite{on_sample_opt}, the authors provide an information-theoretic bound on the number of datapoints required from devices to achieve a fixed generalization error. In \cite{wagle_d2d2022}, the authors leverage embedding exchange to facilitate model alignment. In \cite{semi_decent}, the authors utilize D2D communication to produce weighted averages of local models for aggregation. In \cite{pflcol}, the authors use pairwise similarity between models to enable aggregation of similar models. In contrast, we aim to design a framework for the discovery of D2D graphs which enable improvement in local data diversity, in turn leading to improved convergence \cite{skewscout}, which we detail in Sec. \ref{sec:graph_formation}. D2D-enabled FL has also considered inter-device exchange of model parameters for two purposes: (i) minimizing communication with a central server in semi-decentralized settings \cite{sdc_r43}, where D2D communication is used to mitigate frequent expensive device-to-server communication in deployments at the wireless edge, and (ii)  negating the need for a server entirely in decentralized settings \cite{decentralized_r43}, where D2D model exchanges between devices propagate parameters through the network while aiming to minimize communication costs. In contrast, we utilize D2D communication to exchange local data-related information to improve the speed of convergence by mitigating the bias in local data. Our method can be used as a pretraining method for such decentralized and semi-decentralized methods, as we will illustrate in Sec.~\ref{sec:results}.



\subsubsection{Trust-Aware Inter-Device Cooperation in FL}
While classical federated learning \cite{fedavg} enforces privacy by prohibiting inter-device communication, several recent works have explored the feasibility of private and secure communication between devices.
The approach in \cite{edo} proposes a privacy-preserving energy data sharing system for smart grid users by using inference accuracy as a data importance metric. The method in \cite{vehnet} introduces a model for efficient and secure data sharing for vehicular networks. 
The approach in \cite{d2d_latency_1} improves latency and the method in \cite{d2d_comm_1} improves communication efficiency by caching the exchanged information. 
\cite{wang2024device} enables smarter device-sampling techniques augmented by data offloading between devices. In contrast to the above works, our method is cognizant of inter-device trust, and enables D2D communication that does not violate these trust constraints while improving the data diversity at each device, as described in Sec. \ref{sec:learning_model}.

\subsubsection{Semi-Supervised and Unsupervised FL}
Federated learning in the absence of complete labels is a challenging problem that results in misalignment of local models. A few works have attempted to mitigate this problem. \cite{som} uses self-organizing maps to combat the problem of data heterogeneity at devices, \cite{fedx} uses knowledge distillation coupled with contrastive learning to enable model alignment. \cite{orchestra} orchestrates a globally consistent clustering of devices' data for better generalization. 
\cite{splitnn} uses a one-shot communication between parties to improve model accuracies. 
\cite{semifl} utilizes a small amount of labeled data at the server to improve performance in semi-supervised settings. 
In contrast, for the semi-supervised setting, we propose a label propagation method \cite{lab_prop} to label the unlabeled information and produce efficient D2D communication graphs, while for the unsupervised learning scheme, we find a global subspace which captures the largest collective variance for clustering. 


\vspace{-1.5 mm}

\subsection{Summary of Contributions}

\begin{itemize}
    \item We formulate the optimal D2D graph discovery problem for cooperative data exchange, which aims to maximize the local data diversity at each device in a D2D-enabled FL system while accounting for communication reliability and inter-device trust constraints (Sec.~\ref{sec:problem_formulation}).
    
    \item To solve the resulting NP-hard graph discovery problem, we propose a multi-agent, decentralized RL framework where devices train policies to jointly maximize local rewards for data diversity and reliability and global rewards for network-wide performance while remaining cognizant of trust constraints. This is enabled by the exchange of lightweight, data-opaque representations of local bias between devices  (Sec.~\ref{sec:graph_formation}).
    
    \item We propose several algorithm components to handle supervised and semi-supervised learning tasks (Sec.~\ref{sec:idc_supssp}), including (i) data diversity vectors for lightweight messages and (ii) distance-based metrics to measure diversity of local class information between neighbors. We also incorporate a distributed label propagation method to ensure compatibility with partially labeled datasets.
    
    \item For unsupervised learning settings, we propose a complementary set of algorithm components to account for the lack of labeled data (Sec.~\ref{sec:idc_usp}): (i) joint distributed principal component analysis (PCA) and K-Means clustering for data partitioning, and (ii) locally-computed cluster centroids and covariance matrices serving as lightweight messages, which enables (iii) divergence-based comparisons of data distributions between neighbors through e.g., Kullback Leibler (KL) divergence.
    
    \item We evaluate our method against baselines on five established datasets and for various FL schemes (Sec.~\ref{sec:results}). Our method shows substantial improvements in terms of convergence speed, energy consumption to reach target accuracies, reliability of D2D communication, and robustness against the presence of stragglers. We also demonstrate the ability of our technique to adapt to regression-based learning tasks, dynamic wireless scenarios, alternative data labeling schemes, and other variations.
\end{itemize}

An abridged version of this work appeared in \cite{wagle2023reinforcement}. 
In this extension, we make the following significant additions to [29]: (1) we augment the inter-device cooperation framework to enable compatibility with the semi-supervised settings, introducing distributed label propagation to account for partially labeled datasets, 
(2) as well as with unsupervised settings by developing an implementation of the framework with clustering-based data partitioning and message passing methods as well as data diversity metrics; and (3) we significantly expand upon our experimental results by evaluating our method with additional datasets and a larger set of baselines. We also enhance our experiments to demonstrate the ability of our framework to adapt to variations in the learning task, wireless network, and downstream FL architecture.



\section{System Model and Problem Formulation}
\label{sec:problem_formulation}

\begin{table}[]
    \centering
    \begin{tabular}{|p{1.2cm}|p{6.5cm}|}
        \hline
       Symbol  & Description \\
       \hline
       \multicolumn{2}{|c|}{System Parameters} \\
       \hline
        $N$ & Number of devices \\
        $c_i$ & Device indexed by $i$ \\
        $\mathcal{D}_i$ & Local dataset at device $c_i$ \\
        $L$ & Number of partitions in dataset \\
        $\mathcal{Z}_i^\ell$ & Partition $\ell$ of data at device $c_i$\\
        $\phi_i^t$ & Local model at device $c_i$ at time-step $t$ \\
        $\tau_a$ & Number of time-steps per global aggregation \\
        $\mathbf{P}_D(i,j)$ & Probability of unsuccessful transmission between transmitter $c_j$ and receiver $c_i$ \\
        $\Phi_k$ & Cluster of reliable devices $k$\\
        $\alpha_D$ & Threshold of reliability for clusters of devices\\
        $B_k$ & Inter-cluster communication budget of cluster $\Phi_k$ \\
        $d'_k$ & Inter-cluster communication by cluster $\Phi_k$ \\
        $\mathbf{T}_i$ & Trust matrix associated with device $c_i$ \\
        $\mathbf{A}$ & Adjacency matrix between devices. \\
        $E_i$ & Total number of incoming edges allowed for device $c_i$ \\
        $\mathbf{b}_i[\ell]$ & Diversity threshold for device $i$ for partition $\ell$ \\
        $\mathbf{D}_{j \rightarrow i}$ & Datapoints shared by transmitter $c_j$ with receiver $c_i$ \\
        \hline
        \multicolumn{2}{|c|}{Decentralized Reinforcement Learning} \\
       \hline
        $\pi_i^t$ & Policy at device $c_i$ at time $t$ \\
        $\textbf{s}_i^q$ & $q$-th unique state at device $c_i$ \\
        $\textbf{W}_{i,j}$ & Received Signal Strength (RSS) at device $c_i$ when receiving a signal from device $c_j$ \\
        $\psi_i^R[q,j]$ & Cumulative reward experienced by device $c_i$ when selecting an incoming edge from device $c_j$ when in state $\textbf{s}_i^q$\\
        $\psi_i^C[q,j]$ & Number of times device $c_i$ has selected an incoming edge from device $c_j$ when in state $\textbf{s}_i^q$\\
        $\mathbf{\Omega}_i$ & Experience buffer at device $c_i$\\
        $r_i^L,r_i^G$ & Local and Global rewards at device $c_i$ respectively\\
        \hline
        \multicolumn{2}{|c|}{Learning task-specific} \\
        \hline
        $\mathbf{V}_{j \rightarrow i}[\ell]$ & Number of datapoints from partition $\ell$ that transmitter $c_j$ can share with receiver $c_i$ \\
        $\mathbf{Q}_{j \rightarrow i}[\ell]$ & Number of datapoints from partition $\ell$ requested by receiver $c_i$ from transmitter $c_j$ \\
        $\mathbf{U}_{j \rightarrow i}$ &  Message transmitted from transmitter $c_j$ to receiver $c_i$ \\
        $\mathbf{D}_{i}$ & Initial class distribution vector at device $c_i$ \\
        $\hat{\mathbf{D}_{i}}$ & Final class distribution vector at device $c_i$ \\
        $\bfF$ & Common subspace identified by distributed PCA \\
        $\bfM_i$ & Projection of dataset $\mathcal{D}_i$ on subspace $\bfF$ \\
        $(\mu_i^k,\Sigma_i^k)$ & Centroid and Covariance Matrix respecively of partition $k$ at device $c_i$ \\
        $h_{i,j}^B[p,q]$,         $h_{i,j}^A[p,q]$ & KL divergence between the $p$-th partition at device $c_i$ and the $q$-th partition at device $c_j$ before and after data exchange, respectively.\\
        \hline
    \end{tabular}
    \caption{Key notations used throughout the paper. \vspace{-0.25in}}
    \label{tab:my_label}
\end{table}

\subsection{Network and Learning Models}\label{sec:learning_model}
We consider an FL system over a network of $N$ devices $\calC = \{ c_1, c_2, \dots, c_N\}$, which are regularly aggregated at a central server. Each device $c_i$ has access to a local model $\phi_i^t \in \mathbb{R}^p$ where $p$ is the number of model parameters. $\phi_i^t$ is updated over the training period $t\in[0, T]$ to minimize a local cost function, which is detailed in Section \ref{sec:learning_task}. Local models $\{\phi_i^t\}_{c_i \in \mathcal{C}}$ are aggregated at the server every $\tau_a$ time steps to obtain a global model $\phi_G^t$, which is broadcast to all devices $c_i \in \mathcal{C}$ for local training. Each device also has access to a local dataset $\mathcal{D}_i$, which is divided into $L$ disjoint subsets or partitions $\{\calZ_i^\ell\}_{\ell \in [1,L]}$, such that $\mathcal{D}_i = \cup_{k=1}^L \calZ_i^\ell$. For example, in supervised learning, all datapoints belonging to a certain label are considered a partition. In scenarios where label information is not available, we will perform data partitioning as an additional data processing step, specific to the learning paradigm. These processes are detailed later in Sec.~\ref{sec:idc}.

\subsubsection{Device-to-Device Communication}
We assume that D2D communication can be established among the devices $\mathcal{C}$ in order to exchange a subset of their local datapoints with each other prior to the learning task. Hence, for receiver $c_i$ from a transmitter $c_j$, we define a vector $\mathbf{D}_{j \rightarrow i} \in \mathbb{R}^{L}$ such that $\mathbf{D}_{j \rightarrow i}[\ell]$ defines the number of datapoints from partition $\ell$ that transmitter $c_j$ shares with receiver $c_i$. 

Now, the received signal at the receiving device $c_i$ is influenced by channel conditions between the receiver $c_i$ and the transmitter $c_j$, such as path loss, interference and noise. 
For a system with a predefined transmission power and rate of transmission, these factors are manifested in the probability of unsuccessful transmission $\bfP_D(i,j)$ \cite{fwc}. We assume that $\bfP_D$ can be calculated at each device, and we utilize $\bfP_D$ to design the reward function (Sec. \ref{sec:idc}). While our proposed method is independent of the calculations used to compute $\bfP_D$, in our experiments in Sec. \ref{sec:results}, we use (\ref{eq:prob_drop}) to calculate $\bfP_D$.


\subsubsection{Clusters of Reliable Devices} 
It is important to identify links to remote devices with low probability of communication failure. We therefore partition the devices in $\mathcal{C}$ into $\kappa$ disjoint clusters, given by $\{\Phi_1, \Phi_2, \dots, \Phi_{\kappa}\},$ where each device $c_i$ belongs to a cluster $\Phi_k$, such that devices within a cluster $\Phi_k$ are capable of reliably communicating among themselves. We define a reliable cluster of devices $\Phi_k$ as one in which for all pairs of devices $c_i,c_j \in \Phi_k$, it is required that $\bfP_D(i,j) \leq \alpha_D$, where $\alpha_D$ is a reliability threshold set by the user.
We can now define two forms of D2D communication, namely \textbf{intra-cluster} and \textbf{inter-cluster} communication.


Now, in order to minimize data exchange over unreliable channels (i.e, inter-cluster communication), we define a budget $B_k$ for each cluster $\Phi_k$ such that 
the number of datapoints requested by devices in $\Phi_k$ from devices which are not in $\Phi_k$ can be at most $B_k$, for any $k = 1,2, \dots, \kappa$. Thus, if $\bfD_{j \rightarrow i}$ is the number of datapoints requested by the receiver $c_i \in \Phi_k$, 
\vspace{-1mm}
\begin{align}\label{eq:budget}
     \sum_{c_i \in \Phi_k} \sum_{c_j \notin \Phi_k}  \bfD_{j \rightarrow i} \leq B_k.
\end{align}
\vspace{-2.5mm}
\subsubsection{Transmitter-specific and Data-specific Trust} 


In D2D communication, protection against privacy breaches is necessary such that, devices are prohibited from sharing data with other devices unless the receiver is trusted by the transmitter.
For example, an individual device carrying the data from various disjoint organizations, such as personal data, work emails and medical reports necessitates trust based data exchange. 
Such scenarios require a method of governing trust between various devices at a level of granularity such that only specified receivers can be trusted with specified partitions of data. 

We encode this notion of trust in a transmitter specific trust matrix $\bfT_j$ defined for a given transmitter $c_j$,  denoted by $\bfT_j \in \mathbb{Z}^{N \times L}$, where the rows of $\bfT_j$ correspond to devices in the system and the columns correspond to data partitions. The entries of $\bfT_j$ belong to the set $\{ 0, 1\}$, given by\vspace{-1.5mm}
\begin{align}
    \bfT_j[i,\ell] = \begin{cases}
        1 & \textrm{if $c_j$ trusts $c_i$ with partition $\calZ_j^\ell$}\\
        0 & \textrm{otherwise}.
    \end{cases}
\label{eq:trust_matrix}
\end{align}
This implies that transmitter $c_j$ can share data from partition $\ell$ with receiver $c_j$ only if $\bfT_j[i,\ell]=1$. Thus, in our system model, we do not allow a transmitter $c_j$ to transmit datapoints of partition $\calZ_j^\ell$ to receiver $c_i$ if $\bfT_j[i,\ell] = 0$. Note that $\bfT_j[i,\ell] = 1$ does not imply that $\bfT_i[j,\ell] = 1$.

In practice, domain rules and restrictions will cause different structures to emerge in the trust matrices. For example, from the perspective of the system, if certain devices are deemed to be insecure, then most trust matrices will possess row sparsity, with such devices not receiving any information. Alternatively, if certain types (i.e., partitions) of data are considered sensitive, this will lead to many trust matrices possessing column sparsity, and the range of partitions which can be used to improve data diversity at receiving devices will be restricted. Finally, a given device may only be able to share data from a limited number of partitions with a limited number of devices, leading to a block diagonal pattern within trust matrices, inhibiting the choices of incoming edges and the variety of data that can be received over them. All of these scenarios limit the ability of the system to improve overall local data diversity via data exchange. The development of our graph discovery methodology will make no particular assumptions about the forms of these matrices, thus being applicable to each scenario. For an experimental study of the impact of trust matrix structure on system performance, please refer to Appendix D.



\subsubsection{Learning Task} \label{sec:learning_task}
Once D2D data exchange has been conducted, the local model $\phi_i^t$ at each device is updated at every time step $t$ to achieve a local learning task, as described in Sec. \ref{sec:learning_model}. For the supervised and semi-supervised paradigm, we consider a classification task, where each device $c_i$ has its own local data-set $\calD_i$ which consists of tuples $(d, \ell) \in \mathcal{D}_i$ where $d$ is the feature vector for the datapoint and $\ell$ is the corresponding class. The performance of the local model $\phi_i^t$ is evaluated by a loss function $\mathcal{L}(\phi_i^t, \mathcal{D}_i)$. In the supervised and semi-supervised learning cases, we define the loss function as
\begin{align}
    \mathcal{L}(\phi_i^t, \mathcal{D}_i) = \sum_{(d,\ell) \in \mathcal{D}_i} \texttt{CELoss}(\phi_i^t, d,\ell),
\end{align}
where $\texttt{CELoss}$ is the cross entropy loss between the predicted and ground truth classes. 

Next, we consider the loss function for the unsupervised learning scenario. Unsupervised learning is used as a pretraining methodology to train a function that maps the unlabeled dataset to a latent space with useful properties \cite{simclr}. To do so, we use the contrastive learning \cite{contrastive_learning} framework, which is a form of unsupervised learning where similar datapoints are mapped to be closer in the latent space, while dissimilar datapoints are further apart. For the unsupervised learning paradigm, we define the loss function as
\begin{align}
    \mathcal{L}(\phi_i^t, \mathcal{D}_i) = \sum_{(d,\hat{d}) \in \mathcal{D}_i} \texttt{Triplet}(\phi_i^t, d,\hat{d},\tilde{d}); ~ \tilde{d} = F(d),
\end{align}
where \texttt{Triplet} is the contrastive triplet loss \cite{triplet_loss}, which operates on an anchor $d$, a negative $\hat{d}$ and a positive $\tilde{d}$ and minimizes the Euclidean distance between embeddings of similar datapoints $(d,\tilde{d})$ and maximizes the distance between those of dissimilar datapoints $(d,\hat{d})$. Note that $\hat{d}$ is different from $d$ and chosen randomly. Here, $F$ is a randomly sampled augmentation function such as rotation or blur function. 

In the FL setting, the goal of the system is to learn a global model $\phi_G^*$ such that
\begin{align}\label{eq:global_func}
    \vspace{-2mm}
    \phi_G^* = \underset{\phi \in \mathbb{R}^p}{\arg\min} \sum_{i = 1}^{|\mathcal{C}|} \mathcal{L}(\phi, \mathcal{D}_i).
\end{align}
Thus, the optimal global model is expected to perform the classification task with high accuracy across the global data distribution $\mathcal{D} = \bigcup_{c_i \in \mathcal{C}} \mathcal{D}_i$. 

\subsection{Graph Discovery Problem Formulation}
\label{sec:gp_formulation}
As described in Sec.~\ref{sec:learning_model}, the local dataset at device $c_i$, given by $\mathcal{D}_i$ consists of disjoint subsets $\{\calZ_i^\ell\}_{\ell \in [1,L]}$ such that $\mathcal{D}_i = \cup_{\ell=1}^{L} \calZ_i^\ell$,
for all devices $c_i \in \calC$. Now, the local models $\phi_i^t$ are updated exclusively based on local datasets $\calD_i$, and hence, they are expected to diverge over the training iterations between aggregation \cite{fedavg}, resulting in slow convergence to $\phi_G^*$. Studies such as \cite{robust} have shown that this effect is more pronounced when the local datasets are non-i.i.d. Our aim is to enable faster convergence of $\phi_G^*$ by improving local data diversity through cooperative D2D exchange. 



To measure the improvement in the diversity of the local dataset, we introduce a function $f:\hat{\mathcal{D}} \rightarrow \mathbb{R}$ which calculates local data diversity with the following properties. $f(\cdot)$ operates on a set $\hat{\mathcal{D}}$ -- a subset of the union of all the local datasets at all devices in the system $\cup_{c_i \in \mathcal{C}} \mathcal{D}_i$ -- and produces a scalar which indicates how similar a local data distribution is to the i.i.d-case, which corresponds to a low bias and thus highly diverse local dataset. Such a metric allows us to compare the results of information exchange over a D2D graph. Additionally, if device $c_j$ shares information from partition $\calZ_{j \rightarrow i}^\ell$ with receiver $c_i$, then for any two subsets $\calZ_m \subseteq \calZ_{j \rightarrow i}^\ell$ and $\calZ_{m'} \subseteq \calZ_{j \rightarrow i}^\ell$ such that $|\calZ_m|=|\calZ_{m'}|$, we assume that the chosen diversity metric $f$ satisfies $f(\mathcal{D}_i \cup \calZ_m)$ = $f(\mathcal{D}_i \cup \calZ_{m'})$.
Therefore, our aim is to maximize the value of 
$f(\mathcal{D}_i \cup \calZ_{j \rightarrow i}) -  f(\mathcal{D}_i).$ 
The specific choice of function $f(\cdot)$ will vary for labeled and unlabeled datasets, as we will detail in Sec. \ref{sec:idc}. For example, valid metrics for labeled datasets include the 1-Wasserstein distance \cite{1wasserstein} and the Jensen-Shannon Divergence \cite{jsd}.

Next, we define D2D information exchange as communication across a directed graph $\mathcal{G} = \{\mathcal{C}, \bfA\}$, where $\mathcal{C}$ is the set of devices and $\bfA$ is the adjacency matrix among devices:  
\begin{align*}
\bfA_{ji} = \begin{cases}
    1 , \;\; \textrm{if there is an incoming edge from $c_j$ to $c_i$} \\
    0 , \; \;  \textrm{otherwise}.
\end{cases}
\end{align*}

Specifically, if device $c_j$ shares information from $\calZ_{j \rightarrow i} \subset \mathcal{D}_j$ of its local dataset with device $c_i$, we denote the edge as $\bfA_{ji}=1$. Once $c_i$ receives the information $\calZ_{j \rightarrow i}$, the updated dataset at device $c_i$ can be represented as $\calD_i \cup \calZ_{j \rightarrow i}$.

Now, the D2D exchange of a small number of datapoints that reduce the non-i.i.d skew in local datasets yields significant performance gains in a learning task \cite{skewscout}. However, discovering an optimal D2D graph over the set of all possible graphs is not straightforward due to the additional resource requirements to account for unreliable channels due to network topology. To maximize the impact of D2D exchange without excessive computational overhead being utilized for optimal graph discovery, we restrict every receiver $c_i$ to receive datapoints from at most $E_i$ other remote devices, resulting in at most $E_i$ incoming edges per device over which it receives data. 
Thus, we assume an upper bound $E_i$ on the total number of incoming links each device can maintain, given by $\sum\limits_{j=1}^N\bfA_{ji}\leq E_i,\forall c_i \in \calC.$

\begin{remark}
{\normalfont This upper bound is also motivated by practical federated deployments, e.g., IoT devices \cite{learning_iot}, smartphones \cite{fedphone} and smart wearables \cite{onbody}, each with significant constraints in terms of communication resources. In Section \ref{sec:graph_formation}, we will develop our method assuming $E_i = 1$, i.e., each device has at most one incoming neighbor, and then show how it can be extended to multiple edges via a greedy edge selection scheme.}
\end{remark}

Next, we define a diversity threshold vector $\bfb_i \in \mathbb{R}^L$ for device $c_i$ such that, $c_i$ requires at least $\bfb_i[\ell]$ datapoints for partition $\ell$ to ensure sufficient local data diversity. Thus, during the D2D data exchange, device $c_i$ aims to possess at least $\textbf{b}_i[\ell]$ datapoints from the $\ell$-th partition. If receiver $c_i$ has fewer that $\textbf{b}_i[\ell]$ datapoints, this is achieved by requesting them across the chosen incoming edge, and if transmitter $c_j$ has more that $\textbf{b}_j[\ell]$ datapoints, this is achieved by retaining at least $\textbf{b}_j[\ell]$ datapoints after transmitting across outgoing edges. Note that a transmitter $c_j$ may have multiple outgoing edges. 
In such cases, the selection of the information to be exchanged, $\calZ_{j \rightarrow i}$ is done using deterministic data selection mechanisms, which ensure that all receivers requesting data from the same partition of transmitter $c_j$ are fairly assigned datapoints and transmitter $c_j$ is left with enough datapoints to fulfill the data threshold criterion on $\textbf{b}_j[\ell]$. Thus, the diversity threshold vector $\textbf{b}_i$ has two key purposes. First, it allows the requesting device to calculate the number of datapoints required to ensure that the requested data meaningfully contributes to the data diversity. Second, it ensures that the transmitting device retains a sufficient number of datapoints for itself. These threshold vectors will be directly incorporated into our message passing algorithms described in Sec.~\ref{sec:idc_usp} and Sec.~\ref{sec:idc_supssp}.  

A number of system level constraints such as trust and network reliability also influence the optimal D2D graph. Thus, in addition to maximizing data diversity, the graph discovery method must also maintain inter-device trust between devices as defined in (\ref{eq:trust_matrix}) while exchanging information and maximize the probability of successful communication, as defined in (\ref{eq:prob_drop}).

Thus, we can now describe the process of cooperative D2D information exchange by defining it as an optimal graph discovery problem. We seek to find an optimal graph $\mathcal{G}^*$ such that the communication graph maximizes a chosen data diversity reward metric while abiding by established notions of trust and maximizing reliable communication between devices.
At a high level, we can formulate this problem as
\begin{equation} \label{eq:opt}
\mathcal{G}^{*} \hspace{-1.mm}= \underset{Y, \mathcal{Z} : \mathbf{T}, \mathbf{A}}{\arg\max}\hspace{-1.mm} \sum_{c_i \in \calC} \hspace{-1.mm}\left[ f(\mathcal{D}_i \cup \bfD_{Y(i) \rightarrow i}) -  f(\mathcal{D}_i) - \lambda\mathbf{P}_D(i,Y(i)) \right]
\end{equation}
where $Y$ is the mapping function from $c_i$ to its selected incoming neighbor $c_{Y(i)}$, and $\mathbf{D}_{Y(i)\rightarrow i}$ captures the data transfers from $c_{Y(i)}$ to $c_i$, subject to the trust $\mathbf{T}$ constraints introduced above, and an appropriate definition of the unsuccessful transmission probability matrix $\mathbf{P}_D$.  However, \eqref{eq:opt} is an NP-Hard combinatorial problem, which is infeasible to solve exactly. Thus, we are motivated to consider decentralized RL-based techniques \cite{sutton2018}. The convergence properties of decentralized multi-agent RL methods have been studied extensively in \cite{dcrl}. It has been shown that the training process in such systems, with a shared reward structure between devices and frequent communication between devices, converges asymptotically. As we will see in the following sections, our method satisfies these conditions.  Next, we describe our RL methodology to approximately find $\calG^*$ for the supervised, semi-supervised and unsupervised learning paradigms.

\begin{figure*}[t]
    \centering
    \includegraphics[width=0.99\textwidth]{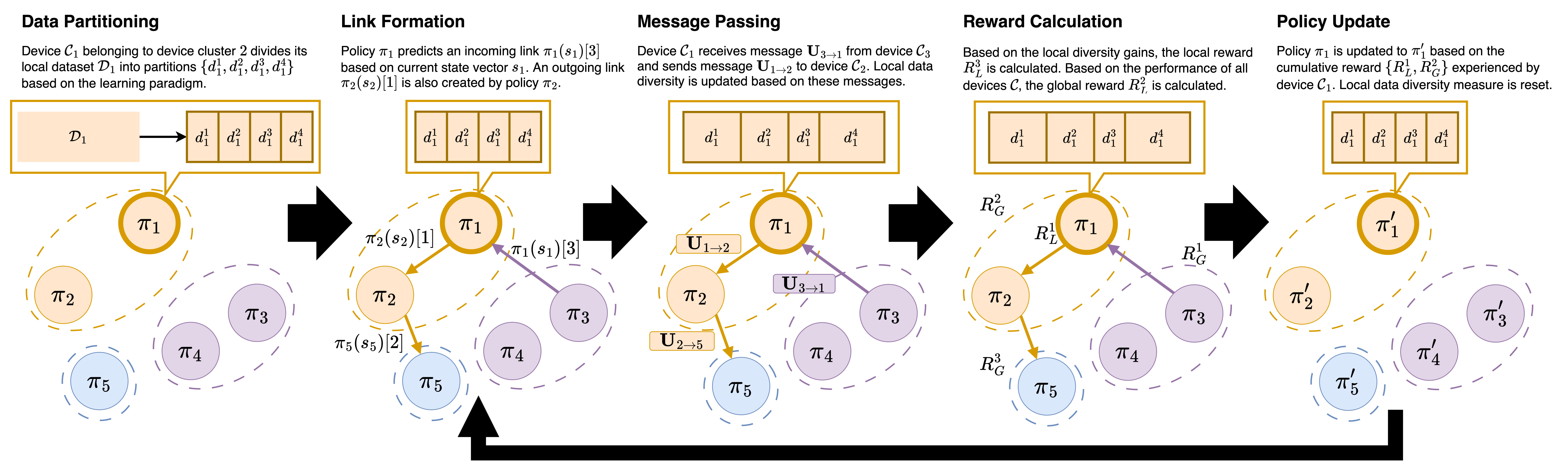}
    \caption{The intelligent graph discovery process iteratively improves local policies in a decentralized manner by updating them such that information exchange over the predicted links maximizes a system-wide performance metric.}
    \label{fig:model}
    \vspace{-4mm}
\end{figure*}

\section{Graph Discovery Framework}
\label{sec:graph_formation}


Our graph discovery methodology consists of five different steps, as illustrated in Fig.~\ref{fig:model}. We start with \textbf{data partitioning}, where each local dataset is divided into partitions containing similar datapoints in terms of the chosen data diversity function $f(\cdot)$ as described in Sec.\ref{sec:gp_formulation}. The notion of similar datapoints is unique to data context, and is discussed further in Sec. \ref{sec:idc}. 
We design the remaining steps based on the Q-Learning framework \cite{sutton2018} to train a policy $\pi_i$ at each device $c_i$. We first conduct \textbf{link formation}, where the set of policies $\{\pi_i \in \mathbb{R}^{A}\}_ {i \in [1,N]}$ predict a set of links over the graph of devices $\calC$. Here, $A=N$ for the supervised and semi-supervised paradigm and $A=\kappa N$ for the unsupervised paradigm, where $\kappa$ is the number of data clusters at each device. These choices are discussed further in methods described in Sec. \ref{sec:idc}. Next, we perform the \textbf{message passing} procedure, which decides the information to be transmitted over the predicted edges as defined in Sec. \ref{sec:idc}.  Next, the \textbf{reward formulation} step calculates the utility of links predicted by the policies $\{\pi_i\}_{i \in [1,N]}$ based on the received information and a reward function. The final step is the \textbf{policy update} which iteratively updates $\{\pi_i\}_{i \in [1,N]}$ based on the experienced rewards, and leads to the discovery of an optimal graph. 

\subsection{Data Partitioning}

For information exchange, not all datapoints at the transmitter are equally important for the receiver. Hence, it is important to distinguish between datapoints in terms of their potential importance for receivers. Thus, for every transmitter $c_j$ we require a method to obtain a set of partitions $\bar{\calZ}_j = \{ \calZ_j^1\ldots \calZ_j^L\}$ such that $\mathcal{D}_j = \cup_{\ell = 1}^L \calZ_j^\ell$ and that any two datapoints belonging to a partition $\calZ_j^\ell$ are similar in terms of the chosen diversity metric $f$ as described in Sec.\ref{sec:gp_formulation}, i.e, For any two datapoints $d_1,d_2 \in \calZ_j^\ell$ and a dataset $\calD'$, $f(\calD' \cup d_1) = f(\calD' \cup d_2)$.

In supervised learning, the data partitions correspond to the labels assigned to the datapoints. For semi-supervised learning, we use a label propagation algorithm, which is explained in Sec. \ref{sec:idc_supssp}, to assign labels to unlabeled datapoints, to enable similar partitions. In the unsupervised case, we utilize a clustering algorithm, which is given in Sec. \ref{sec:idc_usp}, to separate unlabeled datapoints into clusters of similar datapoints.

\subsection{Link Formation}\label{sec:link_formation}
\vspace{-1.mm}
In this step, we first define the state vector of device $c_i$ as $\textbf{s}_i^q = \{\bfW_{i,j} : c_j \in \mathcal{C}\} \in \mathbb{R}^{N}$, where $q$ indexes different instances of the state. Here, $\bfW_{i,j}$ is the received signal strength (RSS) at device $c_i$ while receiving a signal from device $c_j$, which is assumed to be constant through the course of training. We will give a specific model for RSS in Sec.~\ref{sec:setup}. $\bfs_i^q$ is used by the policy $\pi_i$ at device $c_i$ to predict an incoming link. To facilitate this, we define a Q-table \cite{sutton2018} at each device $c_i$ using $\psi_i^R \in \mathbb{R}^{S \times A}$ and $\psi_i^C \in \mathbb{R}^{S \times A}$, where $S$ is the number of unique states experienced by policy $\pi_i$. The Q-table is used to calculate the utility of a predicted link, and will be discussed in Sec. \ref{sec:policy_update}. Here, $\psi_i^R[q,j]$ is the total reward and $\psi_i^C[q,j]$ counts the frequency, respectively, for all times that policy $\pi_i$ selects a link from device $c_j$ when in state $\textbf{s}_i^q$ over the RL training process. Each device $c_i$ predicts an incoming edge from $c_j$ using its local policy $\pi_i$ and $\textbf{s}_i^q$ with probability 
\begin{align}\label{eq:prob_edge}
    \pi_i(\textbf{s}_i^q)[j] = \frac{\exp\left(\frac{\psi_i^R[q,j]}{\psi_i^C[q,j]}\right)}{\sum_{c_k \in \mathcal{C}} \exp\left(\frac{\psi_i^R[q,k]}{\psi_i^C[q,k]}\right)},
\end{align}
where the numerator is proportional to the average reward experienced when selecting a link from $c_j$ when in state $\textbf{s}_i^q$, and the denominator is a normalizing factor that ensures that $0 \leq \pi_i(\textbf{s}_i^q)[j] \leq 1$. Thus, policy $\pi_i$ learns to select links that maximize the total reward. Once links are predicted for all receivers, information is shared across the resulting graph over $\calC$ using the message passing algorithm described next.



\subsection{Message Passing}\label{sec:message_passing}

We use an iterative, exploratory method to discover the optimal graph, which entails some D2D communication overhead \textit{during} the optimal graph discovery phase as well.  
Also, during the graph discovery phase, devices exchange \textit{messages} with other devices, several of which are not selected for exchange of \textit{datapoints}. 
Hence, it is crucial to keep the communication overhead during this phase minimal while keeping datapoint related information private during message passing. To that end, we design a message passing algorithm which shares a compressed form of selected information over a link which is relevant to achieving the graph discovery objective, while not containing any information which can allow the receiver to reproduce specific datapoints at the transmitter. 


In this regard, we design individual message passing algorithms for (i) the supervised/semi-supervised case and (ii) the unsupervised case. In the supervised/semi-supervised case, each transmitter shares the number of datapoints belonging to each label that can be exchanged with the receiver. This message formulation 
is compliant with inter-device notions of trust and is lightweight (an integer vector of size $L$, i.e, $8L$ bits). This algorithm is described in details in Section \ref{sec:idc_supssp}.

For the unsupervised case, we do not have access to ground truth information. Hence, we perform a local clustering procedure to clusters of datapoints at each device. Centroids and covariance matrices of selected clusters are shared between devices to calculate local data diversity, the process of which is explained in Sec. \ref{sec:idc_usp}. This message formulation 
is lightweight (a floating-point vector, a floating-point matrix and an integer, i.e, $32(d + d^2) + 8$ bits, where $d$ is a user-defined parameter). Details of this method are described in Sec. \ref{sec:idc_usp}.



Thus, during the RL training phase, we do not share any datapoints among the devices, and information is only shared if it is (a) permitted by the trust matrix $\mathbf{T}_j$ for transmitter $c_j$ and (b) requested by the receiver $c_i$ from $c_j$. This information is extremely lightweight as compared to typical neural network sizes, resulting in a negligible communication overhead, as we will show in Sec.~\ref{sec:results}. Next, we discuss the formulation of the policy reward, based on the received message.

\subsection{Reward Formulation}\label{sec:reward_formulation}

The overall reward at each device $c_i$ should consider (i) the performance of its local policy $\pi_i$, (ii) performance of other devices $\{c_j \in \calC\}_{j \neq i}$, (iii) reliability of the received signal, as we will define in Sec.\ref{sec:results}, and (iv) the inter-cluster exchange, as defined in (\ref{eq:budget}). The local data diversity, as defined in (\ref{eq:data_diversity}) should be improved to accelerate convergence. For a predicted link between $c_i$ and $c_j$, the probability of failed transmission $\bfP_D(i,j)$ should be low.
We will give a specific model for $\bfP_D(i,j)$ in Sec. \ref{sec:results}.
For a cluster of reliable devices $\Phi_k$, the data shared between clusters must be less than the data budget $B_k$, as defined in (\ref{eq:budget}), to maximize usage of reliable links. 
Trust concerns are handled by the message passing algorithms described, which are described in Sec.~\ref{sec:sup_mp} and \ref{sec:usp_mp} for the supervised/semi-supervised and unsupervised cases, respectively. 

In order to incorporate all of the above metrics, the overall reward must constitute a tradeoff between local and system performance. The reward consists of two components, a {\bf local reward} $r_i^L$ specific to device $c_i$, and a {\bf global reward} $r_k^G$ specific to cluster ${\Phi_k}$. The local reward $r_i^L$ 
captures only the performance of the policy $\pi_i$ at device $c_i$ in terms of data diversity and reliability, while the global reward $r_k^G$ captures the performance of the overall network, ensuring that all devices improve on average while cluster budget constraints are met. To that end, local rewards $r_i^L$ are shared between devices. Budget constraints are found by obtaining the number of datapoints received over inter-cluster links for device $c_{i'} \in {\Phi_k}$ as 
$    d'_k = \sum_{c_i' \in {\Phi_k}} \sum_{\ell=0}^{L}
        |\bfD_{j' \rightarrow i'}[\ell]| $ where $ j' \sim \pi_i({\textbf{s}_i^q})$ and $c_j' \notin {\Phi_k}$.

Thus, the overall reward for a device $c_i \in {\Phi_k}$ is given by $ R_i^k = r_i^L + \gamma \cdot r_k^G$. {We will detail the computations of $r_i^L$ and $r_k^G$ for the supervised/semi-supervised and unsupervised settings in Sec.~\ref{sec:idc_supssp} and \ref{sec:idc_usp}}. The weighting term $\gamma$ governs the importance given to the overall performance of the system. 

\subsection{Policy Update}\label{sec:policy_update}

{We use a decentralized multi-agent Q-Learning algorithm to update the policy $\pi_i$ in a state indexed by $\bfs_i^q$ using an experience buffer $\mathbf{\Omega}_i \in \mathbb{R}^H$, which accumulates the last ${H}$ rewards experienced by $\pi_i$. The reward for $\pi_i$ at device $c_i \in \Phi_k$ when it selects an incoming edge from device $c_j$ when in state $\bfs_i^q$ at RL training step $t$ is denoted by $R_i^k(t)$. Now, we selectively update the policy by biasing the predictions towards actions which result in rewards better than the average reward contained by the buffer. We formalize this as follows:
\begin{align*}
    & \beta = \begin{cases}
        1 - \delta & \textrm{if $R_i^{k}(t) < (\sum_{\hat{t}} \mathbf{\Omega}_i[\hat{t}] / {H})$}\\
        1 & \textrm{otherwise}
    \end{cases},\\
    & \mathbf{\Omega}_i[t'] \leftarrow \beta \cdot R_i^{k}(t), \\
    & \psi_i^R[q,j] = \sum_{\hat{t}=1}^H   \mathbf{\Omega}_i[\hat{t}]\,, \\
    & \psi_i^C[q,j] = \begin{cases}
        t & \textrm{if $t \leq H$}\\
        H & \textrm{otherwise}
    \end{cases};
\end{align*}
where $t' \equiv t \, (\textrm{mod} \; H)$,} and $\delta \in [0,1]$ is a user-defined weight reduction given to rewards that are below the buffer average. The resultant scaling term $\beta$ allows the policy to incentivize actions which improve performance while still learning from suboptimal actions. 

\begin{remark}
\normalfont 
\label{remark:more_edges}
We can apply our graph discovery method sequentially to predict at most $E$ incoming edges for each device in a system of $N$ devices. We perform the graph discovery and data exchange steps $E$ times in succession to discover at most $E$ edges for each device in the system. The sequential method discovers near-optimal graphs with complexity $O(NE)$. In contrast, the complexity of solving the graph discovery problem concurrently is $O(N^E)$, as the action space must accommodate all combinations of edges \cite{sutton2018}. 
\end{remark}


\section{Inter-device Cooperation Methodology}
\label{sec:idc}

\subsection{Supervised and Semi-Supervised Learning}
\label{sec:idc_supssp}

\subsubsection{Data Partitioning} In the supervised learning scenario, 
we partition the dataset $\mathcal{D}_i$ into subsets $\{\calZ_i^\ell\}_{\ell \in L}$, where subset $\calZ_i^k$ contains all local datapoints in $\mathcal{D}_i$ belonging to class $\ell$. 

In the semi-supervised paradigm, we assume that a fraction of the dataset $\calD_i$ is labeled, while the rest are unlabeled, and device $c_i$ contains at least one labeled datapoint for each label in the original, fully labeled local dataset $\calD_i$. 
In order to obtain a fully labeled dataset from the partially labeled dataset, we propose a distributed label propagation method which assigns labels to unlabeled datapoints in each local dataset. This consists of two steps, where in the first step, we perform distributed PCA \cite{distpca} on local datasets to identify a common subspace without exchanging data information. This subspace captures the directions of highest variance of the complete dataset and avoids the ``curse of dimensionality" \cite{curse_of_dimensionality} for the next step. Next, in the second step, we perform the label propagation algorithm \cite{lab_prop} on the projections of each local dataset on the common subspace, which iteratively generates and updates the labels of each unlabeled datapoint based on the labels of proximal datapoints. Thus, after we perform label propagation on each local dataset $\mathcal{D}_i$, we can partition  $\mathcal{D}_i$ as $\{\calZ_i^\ell\}_{\ell \in L}$, where $\calZ_i^\ell$ contains all datapoints in $\mathcal{D}_i$ with label $\ell$. 

Now, using the label information of the local data, we define the class-distribution vector at device $c_i$ as $\bfD_i \in \mathbb{R}^{L}$, where $L$ is the total number of classes in global dataset $\mathcal{D}$ and $\ell$-th entry of $\bfD_i$ is the number of local datapoints of class $\ell$ available in device $c_i$. 
Now, we take into account the skew of classes across devices by first defining a diversity threshold $\hat{L}$, which is set by the user. We ensure that each device $c_i$ has at least $\hat{L}$ classes available in their local dataset after D2D exchange by imposing the following constraint:
\begin{align}\label{eq:data_diversity}
    \left( \sum_{\ell = 1}^L \mathbbm{1}_{\bfD_i[\ell] \geq \bfb_i[\ell]} \right) \geq \hat{L} .
    \vspace{-2.mm}
\end{align} 
Here, $\bfb_i$ is the threshold vector described in Sec. \ref{sec:problem_formulation}. $\bfb_i$ can be user defined, as different scenarios may limit the number of datapoints that can be shared over wireless channels. 

As the data partitions for semi-supervised learning are now labeled and data diversity vectors have been calculated, we can use identical message passing mechanisms and data diversity metrics in both the supervised and semi-supervised paradigm.

\subsubsection{Message Passing}
\label{sec:sup_mp}

\begin{algorithm}[t]
\caption{D2D Message Passing for Supervised or Semi-Supervised Scenarios} \label{algo:message_passing}
\begin{algorithmic}[1]
    \State \textbf{Given :} Receiver node $c_i$, Selected transmitter node $c_j$, current state $s$, policy $\pi$.
    \State Transmitter $c_j$ communicates the labels that can be shared with receiver $c_i$ as $\bfV_{j \rightarrow i}$ using (\ref{eq:avl_at_tx}).
    \State Receiver $c_i$ finds the data diversity $\bfD_i$ according to (\ref{eq:data_diversity}).
    \State Receiver $c_i$ finds the required data vector $\bfQ_{j \rightarrow i}$ using (\ref{eq:req_at_dst}).
    \State Transmitter $c_j$ updates message $\bfU_{j \rightarrow i}$ using (\ref{eq:tx_buffer}) and transmits them to receiver $c_i$.
\end{algorithmic}
\end{algorithm}

 Let $\mathcal{N}_j$ be the set of devices requesting datapoints from transmitter $c_j$ after the link formation step. 
 Device $c_j$ shares the indices of labels that can be shared with each device $c_i \in \mathcal{N}_j$ as a vector $\bfV_{j \rightarrow i} \in \mathbb{R}^L$.
 Now, we calculate the vector $\bfV_{j \rightarrow i}[\ell]$ as follows
\begin{align}\label{eq:avl_at_tx}
\bfV_{j \rightarrow i}[\ell] = \begin{cases}
    1 &\; \texttt{if} \; \bfT_j[i,\ell] = 1 \; \texttt{\&} \; c_i \in \mathcal{N}_j \; \texttt{\&} \; \bfD_{j}[\ell] > \bfb_j[\ell]\\
    0 & \; \texttt{otherwise}.
\end{cases}
\end{align}
The above equation ensures that transmitter $c_j$ only shares those datapoints that are allowed by trust matrix $\bfT_j$ while satisfying data diversity requirements. 

Upon receiving $\bfV_{j \rightarrow i}$, receiver $c_i$ forms a requirement vector $\bfQ_{j \rightarrow i} \in \mathbb{R}^{L}$, where $\bfQ_{j \rightarrow i}[\ell]$ is the number of datapoints of class $\ell$ requested by $c_i$ from $c_j$ and is calculated as follows:
\vspace{-1mm}
\begin{align}\label{eq:req_at_dst}
\bfQ_{j \rightarrow i}[\ell] = \begin{cases}
        \bfR_i[\ell] , &\texttt{if } \bfR_i[\ell] > 0 \texttt{ \& } \bfV_{j \rightarrow i}[\ell]=1\\
        0 , &\texttt{else }
        \end{cases},\vspace{-2.mm}
\end{align}
where $\bfR_i[\ell] = \bfb_{i}[\ell] - \bfD_i[\ell]$ denotes the number of datapoints of label $\ell$ required by receiver $c_i$ to meet the diversity threshold. 
The requirement vector $\bfQ_{j \rightarrow i}$ is then shared with transmitter $c_j$. Based on $\bfQ_{j \rightarrow i}$, $c_j$ selects datapoints of class $\ell$ from $\mathcal{D}_j$ for all classes $\ell \in \{1, 2, \dots, L\}$ and forms a message $\bfU_{j \rightarrow i} \in \mathbb{R}^L$. 
The message $\bfU_{j \rightarrow i} \in \mathbb{R}^L$ contains the number of datapoints that $c_j$ is \textit{actually} able to share. Note that this may differ significantly from $\bfV_{j \rightarrow i}$ due to the different demands $\bfQ_{j \rightarrow i'}$ made by all $i' \in \mathcal{N}_j$ devices. If the total demand is higher than what $c_j$ can afford to transmit, datapoints are sent based on relative demand from each receiver $c_i\in \mathcal{N}_j$. We calculate message $\bfU_{j \rightarrow i}$ as follows:
\vspace{-1mm}
\begin{align}\label{eq:tx_buffer}
\bfU_{j \rightarrow i}[\ell] = \begin{cases}
    \bfQ_{j \rightarrow i}[\ell] , \;\; \texttt{if} \sum_{c_{i'} \in \mathcal{N}_j} \bfQ_{j \rightarrow i'}[\ell] \leq \bfD_{j}[\ell] - \bfb_j[\ell] \\
    \frac{\bfQ_{j \rightarrow i}[\ell]}{\sum\limits_{c_{i'} \in \mathcal{N}_j} \bfQ_{j \rightarrow i'}[\ell]} \cdot (\bfD_{j}[\ell] - \bfb_{j}[\ell]) , \;   \texttt{else.}
\end{cases}\vspace*{-5mm}
\end{align}
Now, as message $\bfU_{j \rightarrow i}$ may drop packets with probability $\bfP_D(i,j)$ as per (\ref{eq:prob_drop}), receiver $c_i$ receives a buffer $\tilde{\bfD}_{j \rightarrow i}$, such that $\tilde{\bfD}_{j \rightarrow i}[\ell] \leq \bfU_{j \rightarrow i}[\ell]~\forall~\ell \in L$ and forms an updated class distribution vector $\hat{\bfD}_i$ as
\begin{align}\label{eq:final_diversity_vector}
\hat{\bfD}_i[\ell] = \bfD_i[\ell] + \tilde{\bfD}_{j \rightarrow i}[\ell] - \hspace*{-2mm}\sum_{c_k \in \mathcal{N}_i} \bfU_{i \rightarrow k}[\ell]. 
\end{align}
In our simulations, we model the expected number of received datapoints $\tilde{\bfD}_{j \rightarrow i}$ as $\tilde{\bfD}_{j \rightarrow i}[\ell] = [1-\bfP_D(i,j)]\bfU_{j \rightarrow i}[\ell]$.\footnote{{Note that in the exchange~\eqref{eq:final_diversity_vector}, we assume that transmitter $c_i$ will not retain its local datapoints $\bfU_{i \rightarrow k}[\ell]$ that it has transmitted. This limits the amount of data which must be exchanged to result in local distributions that are closer to i.i.d. Similar approaches are seen in D2D-enabled FL approaches that consider data discarding, e.g.,~\cite{wang2021network}.}}

The message passing algorithm is outlined in Alg. \ref{algo:message_passing}. For a motivational example to illustrate the message passing algorithm, see Appendix A in the supplemental material.

\subsubsection{Data Diversity Reward Metric}

Now, we use the updated class distribution vectors $\hat{\bfD_i}$ to formulate a suitable reward. This enables the policies to learn device-specific requirements via a local reward, while also optimizing the system-wide metrics via a global reward. We start with the local reward.

In order to account for the data diversity requirement (\ref{eq:data_diversity}), 
we first define a score function $g:(\mathbb{R}^L,\mathbb{R}^L) \rightarrow \mathbb{R}$, which maps a diversity vector $\bfD_i$ and a set of threshold values $\bfb_i$ as \vspace*{-1mm}
\begin{align*}\label{eq:diversity_function}
    g(\bfD_i,\bfb_i) = \begin{cases}
         \texttt{Wass}(\bfD_i,\hat{\bfD}_i), & \texttt{if} \; \left(\sum_{\ell = 1}^L \mathbbm{1}_{\hat{\bfD}_i[\ell] \geq \bfb_i[\ell]} \right) \geq \hat{L}\\
         0, & \texttt{otherwise}.
     \end{cases} \vspace*{-7mm}
\end{align*}
Here, \texttt{Wass}$(X_1,X_2)$ is the 1-Wasserstein distance \cite{1wasserstein} between discrete probability distributions $X_1$ and $X_2$. 
The score function $g$ ensures that the predicted links satisfy the data diversity requirement in \eqref{eq:data_diversity}, by only returning rewards if the condition is met. Thus, we define the local reward as 

\vspace{-1mm}
\begin{equation}
\label{eq:sup_local_reward}
    r_i^L = \underbrace{\alpha_1 \cdot g(\hat{\bfD}_i,\bfb_i)}_{\text{Data Diversity}} - \underbrace{\alpha_2 \cdot  (\bfP_D(i,j))}_{\text{Reliability Maximization}} , j \sim \pi_i({\textbf{s}_i^q}).    
\end{equation}

    \vspace{-0.1 cm}


We now define the global reward for a cluster of reliable devices ${\Phi_k}$ as
\vspace*{-2mm}
\begin{align}
\label{eq:sup_global_reward}
    r_k^G = \underbrace{\sum_{c_i \in \calC} \frac{r_i^L}{N}}_{\text{System Performance}} + \underbrace{\alpha_3 \cdot ({B_k} - {d'_k})}_{\text{Cluster Budget}},
    \vspace*{-1mm}
\end{align}
where the cluster budget for device cluster ${\Phi_k}$ given by $d'_k$, limits communication over unreliable links based on an allocated data budget ${B_k}$ as described in Sec. \ref{sec:reward_formulation}.


\subsection{Unsupervised Learning}
\label{sec:idc_usp}

\begin{figure*}
    \centering
    \includegraphics[width=0.98\textwidth]{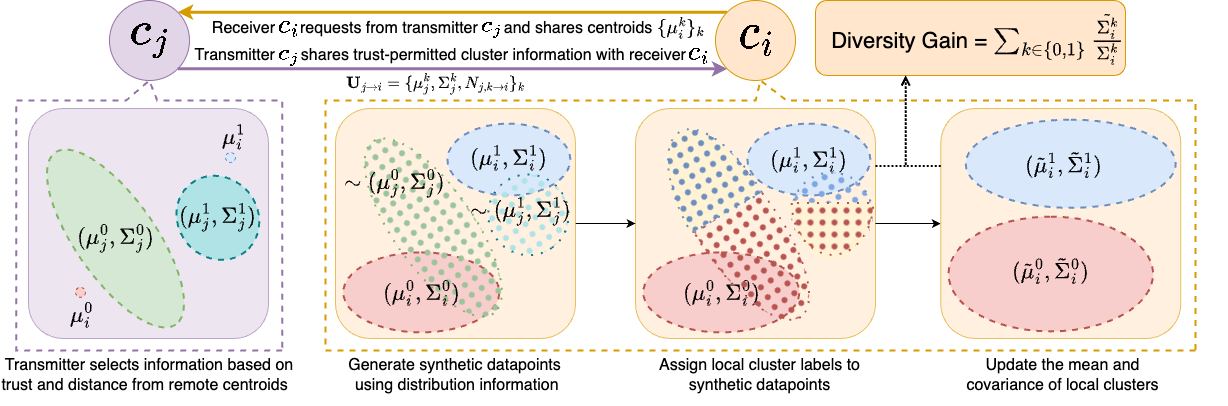}
    \caption{Message passing and data diversity calculation in the unsupervised learning paradigm captures the gain in diversity through the increase in the size of the local clusters after data exchange in terms of the trace of their covariance matrices. During the graph discovery phase, this process involves only the exchange of cluster distribution information without exposing the local datapoints at either device.}
    \label{fig:usp_diversity}
    \vspace{-2mm}
\end{figure*}
\subsubsection{Data Partitioning} 
\label{sec:usp_dp}
In order to partition unlabeled data, we perform distributed dimensionality reduction followed by a clustering algorithm to produce partitions containing similar datapoints. 
First, we reduce the dimensionality of data at each local dataset $\{\calD_i\}_{i=1:N}$ using distributed PCA \cite{distpca}, which allows the system to identify a common subspace denoted by $\mathbf{F} \in \mathbb{R}^{D \times d}$, where $D$ is the feature size of each datapoint, and $d$ is a user-defined parameter such that $d < D$. The subspace $\mathbf{F}$ captures the intrinsic structure of the data by retaining the directions of highest variance within the complete dataset without explicit exchange of data between devices. We then calculate the projections of each local dataset $\mathcal{D}_i$ on the calculated subspace as $\mathbf{M}_i = \mathcal{D}_i \cdot \mathbf{F}$. This process mitigates the ``curse of dimensionality" \cite{curse_of_dimensionality} which adversely affects the clustering process discussed next. We segregate the local data at each device into clusters by performing K-means clustering 
on $\mathbf{M}_i$ to obtain $L$ clusters ${\{\calZ_i^1, \calZ_i^2, \dots, \calZ_i^L\}}$, thus ensuring that all datapoints within a cluster are structurally similar. Thus, we can obtain partitions of unlabeled data at every device as the clusters resulting from the K-Means clustering process. 

\subsubsection{Message Passing} 
\label{sec:usp_mp}
Compared to labeled datasets, where messages can be defined in terms of the number of datapoints belonging to each label, the same cannot be done for unlabeled dataset, as cluster assignments by the K-Means process given in Sec. \ref{sec:usp_dp} are not shared across devices due to the distributed nature of clustering. Hence, we utilize the centroid and covariance information of each cluster to estimate data diversity, as we will describe in Sec.~\ref{sec:usp_div}. To that end, we design a novel message passing method which selects data based on the proximity of remote centroids. Our method is independent of the discrepancy between cluster assignments across devices, making it applicable for the unsupervised federated learning scenario. We define the method as follows.

The receiver $c_i$ requests $\bfQ_{j \rightarrow i}$ datapoints from transmitter $c_i$ and also shares its local centroids $\{\mu_i^1, \mu_i^2, \dots, \mu_i^L\}$ obtained through the K-Means clustering process described in Sec. \ref{sec:usp_dp}. Transmitter $c_j$ now calculates the number of datapoints to share from each cluster $\ell$ as $\tilde{\bfQ}_{j,\ell\rightarrow i}$ as
\begin{align}
\label{eq:usp_mp}
    \tilde{\bfQ}_{j,\ell\rightarrow i} = \bfQ_{j \rightarrow i} \cdot \frac{\sum_{\ell'=1}^L ||\mu_i^{\ell'} - \mu_j^\ell||_2}{\sum_{\hat{\ell}=1}^L \sum_{\ell'=1}^L ||\mu_i^{\ell'} - \mu_j^{\hat{\ell}} ||_2}.
\end{align}

By using (\ref{eq:usp_mp}), to select the number of datapoints from each cluster, the transmitter shares data that is further away from the receiver centroids $\{\mu_i^1, \ldots, \mu_i^L\}$, and is thus, less likely to be present in the local dataset at the receiver. 
The transmitter then calculates the number of datapoints that are available to send from each cluster ${\bfV}_{j,\ell\rightarrow i}$, based on the number of devices requesting datapoints using (\ref{eq:avl_at_tx}). The number of datapoints from each cluster that are finally chosen to be sent is given by ${\bfD}_{j,\ell \rightarrow i} = \min ({\bfV}_{j,\ell\rightarrow i},\tilde{\bfQ}_{j,\ell\rightarrow i})$. Thus, the message sent from transmitter $c_j$ to receiver $c_i$ is $\mathbf{U}_{j \rightarrow i} = \{\mu_j^\ell, \Sigma_j^\ell,\bfD_{j,\ell \rightarrow i}\}_{\ell=1}^L$. The corresponding algorithm is described in Alg. \ref{algo:message_passing_usp}.

\begin{algorithm}[t]
\caption{Unsupervised D2D Message Passing and Local Reward Calculation 
} \label{algo:message_passing_usp}
\begin{algorithmic}[1]
\State \textbf{Given :} Receiver $c_i$, Transmitter $c_j$, receiver policy $\pi_i$.
\State Receiver $c_i$ selects transmitter $c_j$ using policy $\pi_i$.
\State Receiver $c_i$ shares local centroids $\{\mu_i^\ell\}_{\ell=1}^L$ with $c_j$.
\State Transmitter $c_j$ allots the number of datapoints to be selected from each local cluster $k$ as $\tilde{\bfQ}_{j,\ell \rightarrow i}$ using (\ref{eq:usp_mp}).
\State Transmitter $c_j$ calculates the number of datapoints available for exchange as ${\bfV}_{j,\ell \rightarrow i}$ using (\ref{eq:avl_at_tx}).
\State Transmitter $c_j$ calculates number of datapoints that can be sent as $\bfD_{j,\ell \rightarrow i} = \min({\bfV}_{j,\ell \rightarrow i},\tilde{\bfQ}_{j,\ell \rightarrow i})$.
\State Transmitter $c_j$ shares centroid $\mu_j^\ell$ and covariance matrix $\Sigma_j^\ell$ and number of datapoints that can be shared with receiver $c_i$ as $\mathbf{U}_{j \rightarrow i} = \{\mu_j^\ell, \Sigma_j^\ell, \bfD_{j,\ell \rightarrow i}\}_{\ell = 1}^L$.
\State Receiver $c_i$ generates probability distribution for each remote cluster $\ell$ defined by $(\mu_i^\ell,\Sigma_i^\ell)$ and samples $\bfD_{j,\ell \rightarrow i}$ datapoints from it, given by $\tilde{\mathcal{D}}_{j \rightarrow i}$. 
\State Receiver $c_i$ updates its local clusters ${\{\calZ_i^\ell\}_{\ell=1:L}}$ with generated data $\tilde{\mathcal{D}}_{j \rightarrow i}$ to obtain new clusters ${\{\tilde{\calZ}_i^\ell\}_{\ell=1:L}}$, with mean $\tilde{\mu_i^\ell}$ and variance $\tilde{\Sigma_i^\ell}$ for each cluster ${\tilde{\calZ}_i^\ell}$.
\State Receiver $c_i$ calculates local diversity gains as $\sum_{\ell=1}^L \frac{\Tr{(\tilde{\Sigma_i^\ell}})}{\Tr(\Sigma_i^\ell)}$.
\end{algorithmic}
\end{algorithm}

\begin{table*}[t]
\small
\begin{tabular}{|l|l|l|l|l|}
\hline
Data               & Paradigm        & Partitioning                        & Message Passing                 & Diversity Metric                   \\ \hline
Unlabeled         & Unsupervised    & Distributed PCA + K-Means           & Cluster Centroid and Covariance & Cov. Trace and KL Div. \\ 
Partially Labeled & Semi-Supervised & Distributed PCA + Label Prop. & Label Diversity Vectors         & 1-Wasserstein Distance             \\ 
Labeled           & Supervised      & None                                & Label Diversity Vectors         & 1-Wasserstein Distance             \\ \hline
\end{tabular}
\caption{The inter-device cooperation framework adapts to all learning paradigms by defining separate methods for data partitioning, message passing and diversity calculations, which enable devices to identify important datapoints, pass lightweight information and improve the utility of a D2D graph. 
}
\label{tab:idc_methods}
\vspace{-4mm}
\end{table*}


\subsubsection{Data Diversity Reward Metric} 
\label{sec:usp_div}
Assuming an incoming edge to receiver $c_i$ from transmitter $c_j$, 
the local reward at device $c_i$ should reflect the benefit gained by receiving data from $c_j$. Device $c_i$ calculates the local reward as follows. 

\begin{enumerate}
    \item Device $c_i$ defines a Gaussian distribution $\calN(\mu_j^\ell,\Sigma_j^\ell)$ and samples $\bfD_{j,\ell \rightarrow i}$ datapoints for each remote cluster $\ell$. The set of all such sampled datapoints is given by $\tilde{\mathcal{D}}_{j \rightarrow i}$.
    \item Device $c_i$ updates its local clusters with generated data $\tilde{\mathcal{D}}_{j \rightarrow i}$ to obtain new clusters ${\{\tilde{\calZ}_i^1,\ldots,\tilde{\calZ}_i^L\}}$, with mean and variance $\tilde{\mu_i^\ell}$ and $\tilde{\Sigma_i^\ell}$ respectively for each cluster ${\tilde{\calZ}_i^\ell}$. 
    \item Device $c_i$ finds local diversity gains given by $\sum_{\ell=1}^L \frac{\Tr{(\tilde{\Sigma_i^\ell}})}{\Tr(\Sigma_i^\ell)} $. 
\end{enumerate} 

In the local diversity metric, the trace of the covariance matrix $\Tr(\Sigma_i^\ell)$ is proportional to the total variation of the cluster $\ell$ \cite{variation}. Thus, the sum of the traces of covariance matrices $\{\Sigma_i^\ell\}_{\ell=1:L}$ indicates the volume of latent space occupied by the data at device $c_i$. 
Note that the denominator $\Tr(\Sigma_i^\ell)$ is not a function of link selection, and hence local diversity increases as the numerator, that is, the sum of the traces after data exchange increases. 
Thus, the proposed metric is proportional to the increase in the volume of latent space covered by data at $c_i$ after receiving $\tilde{\mathcal{D}}_{j \rightarrow i}$. The message passing and subsequent data diversity calculation processes are illustrated in Fig. \ref{fig:usp_diversity}. We now calculate the local reward as
\begin{align}
\label{eq:usp_local_reward}
    r_i^L = \underbrace{\alpha_1 \cdot \sum_{\ell=1}^L \frac{\Tr{(\tilde{\Sigma_i^\ell}})}{\Tr(\Sigma_i^\ell)} }_{\text{Data Diversity Gains}} - \underbrace{\alpha_2 \cdot  (\mathbf{P}_D(i,j))}_{\text{Reliability Maximization}} , j \sim \pi_i({\textbf{s}_i^q}).
\end{align}

We also calculate a global reward which reflects the overall gain of the system. The global reward is calculated as follows:

\begin{enumerate}
    \item For every device $c_i$ we define Gaussian distributions given by $\mathcal{N}(\mu_i^\ell,\Sigma_i^\ell)$ and $\mathcal{N}(\tilde{\mu}_i^\ell,\tilde{\Sigma}_i^\ell)$.
    \item For each pair of devices $c_i,c_j$ for each cluster {$\calZ_i^\ell,\calZ_j^{\ell'}$}, we calculate the KL Divergence before and after exchange as ${h_{i,j}^B[\ell,\ell'] = \texttt{KL}(\mathcal{N}(\mu_i^\ell,\Sigma_i^\ell)\parallel\mathcal{N}(\mu_j^{\ell'},\Sigma_j^{\ell'}))}$ and ${h_{i,j}^A[\ell,\ell'] = \texttt{KL}(\mathcal{N}(\tilde{\mu}_i^\ell,\tilde{\Sigma}_i^\ell)\parallel \mathcal{N}(\tilde{\mu}_j^{\ell'},\tilde{\Sigma}_j^{\ell'}))}$ respectively. 
    \item The system agreement is then given by $\sum_{i,j} \sum_{\ell,\ell'} {\frac{h_{i,j}^B[\ell,\ell']}{h_{i,j}^A[\ell,\ell']}}$.
\end{enumerate}

For a motivating example to demonstrate the efficacy of our designed diversity metric, see Appendix B.

Note that the numerator ${h_{i,j}^B[\ell,\ell']}$ is not a function of the link selection, hence system agreement increases as the denominator, that is, the post-exchange KL Divergence between clusters decreases, thus encouraging the formation of similar datasets across devices, thereby making the distributions more i.i.d.

We now define the global reward for device cluster ${\Phi_k}$ as 
\begin{equation}
\label{eq:usp_global_reward}
    r_k^G = \underbrace{\sum_{i,j} \sum_{\ell,\ell'} ({\frac{h_{i,j}^B[\ell,\ell']}{h_{i,j}^A[\ell,\ell']}})}_{\text{System Agreement}} + \underbrace{\alpha_3 \cdot ({B_k} - {d'_k})}_{\text{Cluster Budget}}.
\end{equation}


Thus, to summarize, we establish the overall inter-device cooperation scheme and graph discovery framework for labeled, partially labeled and unlabeled datasets. For each learning paradigm, we define data partitioning methods which separate local datasets into subsets consisting of similar datapoints, we design lightweight messages which convey information critical for data diversity calculation while not exposing datapoint related information and we calculate local data diversity based on these messages which enable us to quantify the utility of selected links. This method promotes the discovery of graphs which maximize local data diversity, while maintaining inter-device trust constraints and reliable inter-device communication. A summary of these methods is given in Table \ref{tab:idc_methods}.


\begin{figure*}[t!]
    \centering
    \begin{subfigure}[t]{0.24\textwidth}
        \includegraphics[width=0.99\linewidth,height=3.5cm]{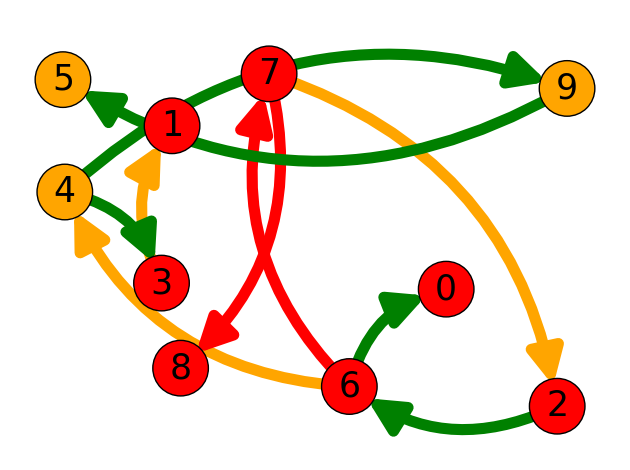}   
        \caption{}\label{fig:gg_unif}
    \end{subfigure}
    \begin{subfigure}[t]{0.24\textwidth}
        \includegraphics[width=0.99\linewidth,height=3.5cm]{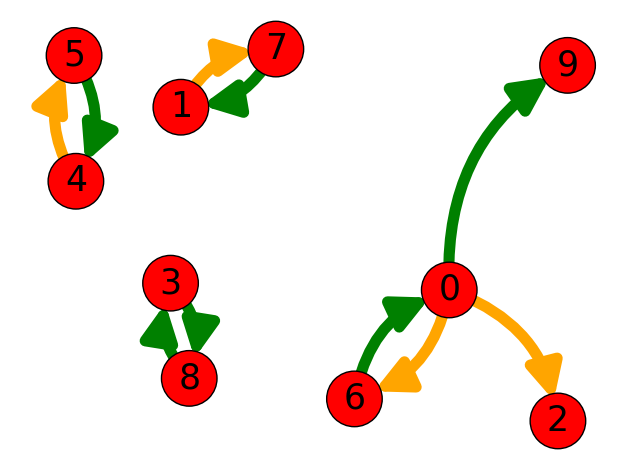}    
        \caption{}\label{fig:gg_unif}
    \end{subfigure}
    \begin{subfigure}[t]{0.24\textwidth}
        \includegraphics[width=0.99\linewidth,height=3.5cm]{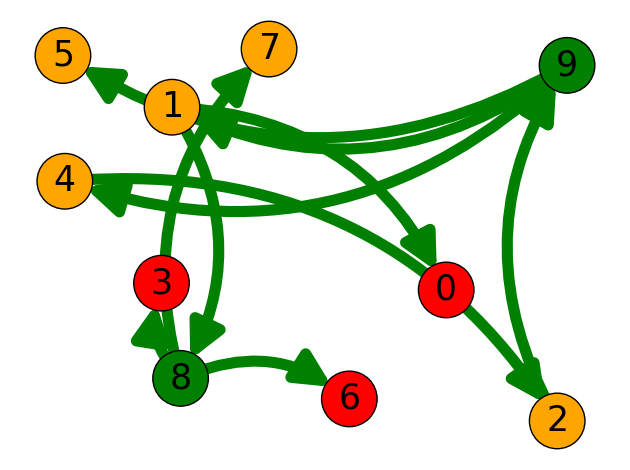}  
        \caption{}\label{fig:gg_unif}
    \end{subfigure}
    \begin{subfigure}[t]{0.24\textwidth}
        \includegraphics[width=0.99\linewidth,height=3.5cm]{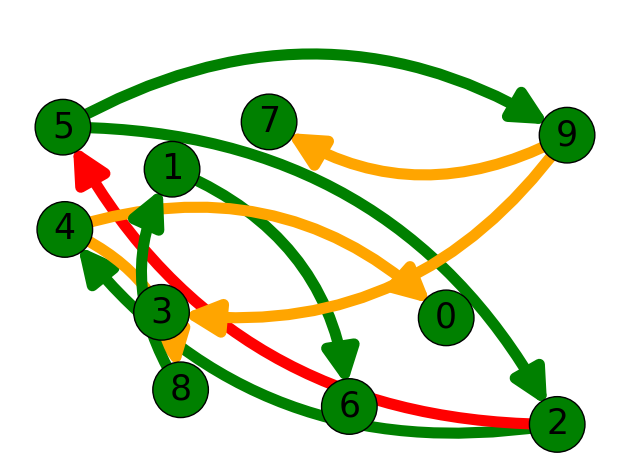}  
        \caption{}\label{fig:gg_unif}
    \end{subfigure}
    
    \caption{D2D graphs generated using (a) Uniform ER, (b) Closest, (c) Most Trusted baseline methods and (d) our method. At each device (node), the number of labels for which the diversity threshold is satisfied is indicated by red, orange and green nodes for 3, 4 and 5 or more labels respectively. For each edge, the number of labels that can be exchanged between devices connected by it (trusted) is indicated by red for less than 3 labels, orange for between 3 and 5 labels, and green edges for 5 or more labels respectively. Our method finds a graph which maximizes diversity, and is significantly different from baselines, indicating that discovering a graph which maximizes diversity is non-trivial.}
    \label{fig:gengraph}
    \vspace{-2mm}
\end{figure*}

\begin{figure*}[h!]	
	\centering
	\begin{subfigure}[t]{0.24\textwidth}
		\centering
		\includegraphics[width=0.99\linewidth,height=3.5cm]{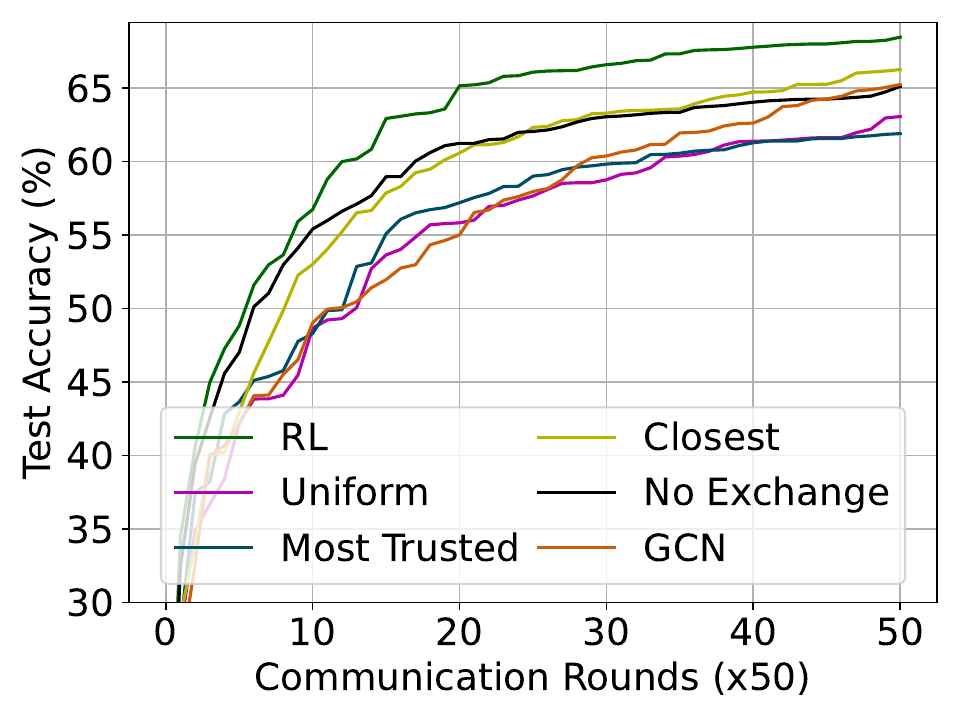}
            \vspace{-7mm}
		\caption{}\label{fig:2a}		
	\end{subfigure}
        \begin{subfigure}[t]{0.24\textwidth}
		\centering
            \includegraphics[width=0.99\linewidth,height=3.5cm]{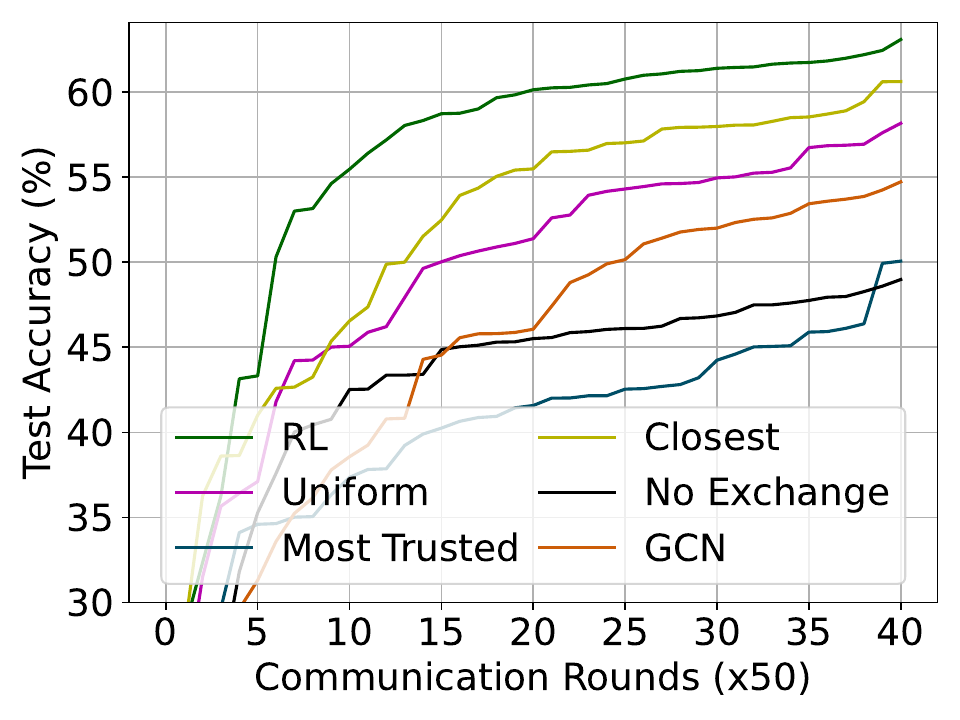}
            \vspace{-7mm}
		\caption{}\label{fig:2f}		
	\end{subfigure}
        \begin{subfigure}[t]{0.24\textwidth}
		\centering
		\includegraphics[width=0.99\linewidth,height=3.5cm]{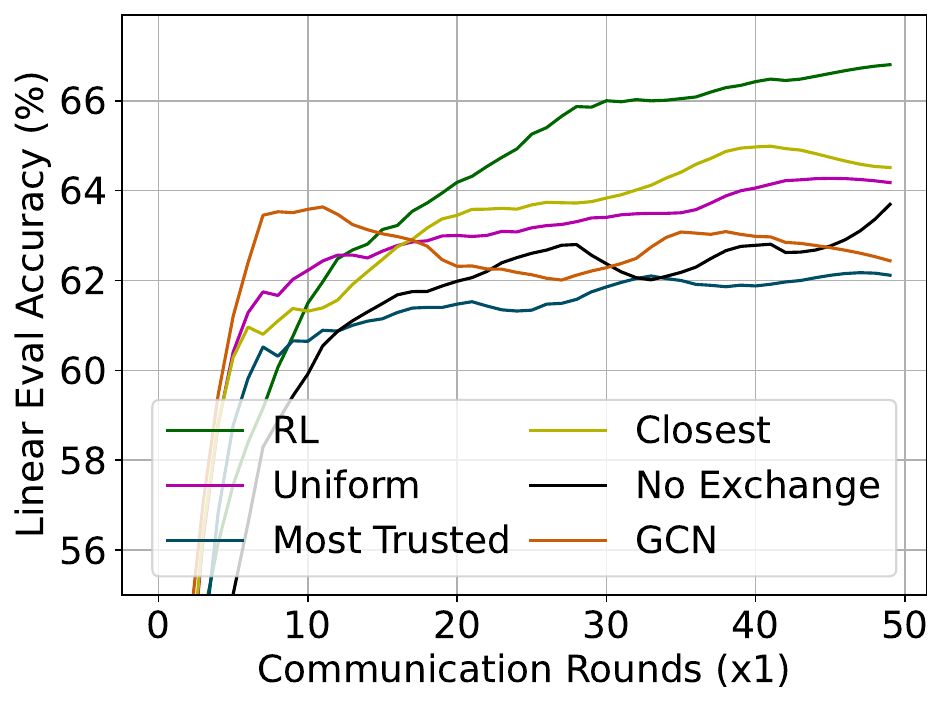}
            \vspace{-7mm}
		\caption{}\label{fig:2d}		
	\end{subfigure}
        \begin{subfigure}[t]{0.24\textwidth}
		\centering
		\includegraphics[width=0.99\linewidth,height=3.5cm]{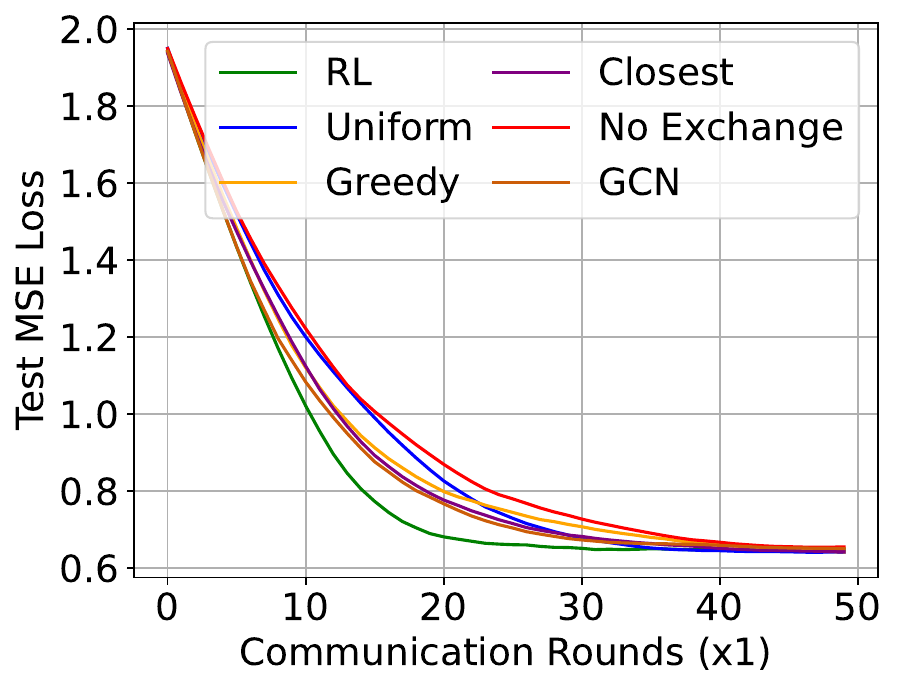}
            \vspace{-7mm}
		\caption{}\label{fig:2h}		
	\end{subfigure}
 
	\caption{Our method cooperatively discovers optimal D2D communication graphs and significantly improves model training convergence performance over baselines. Here, the training process is shown for CIFAR-10 dataset in the supervised setting (Fig. a), RadioML dataset in the semi-supervised setting (Fig. b), the FMNIST dataset for the unsupervised setting (Fig. c), and the California Housing dataset for a regression task, where we evaluate our method using the test mean squared error loss. (Fig. d).} 
 
 \label{fig:2M}
 \vspace{-2.7mm}
\end{figure*}

\section{Simulation Results and Discussion}\label{sec:results}



\subsection{Experimental Setup}
\label{sec:setup}
A detailed description of our setup can be found in Appendix C in the supplementary material.
To summarize, in our experiments, for the supervised and semi-supervised FL scenarios, we employ the CIFAR-10 and SVHN datasets, as well as the RadioML \cite{radioml} signal classification dataset. By default, we consider a network of $N = 25$ devices. We use the Alexnet \cite{alexnet} model, and allow at most one incoming edge.

For the unsupervised federated learning scenario, we use the Fashion-MNIST (FMNIST) and USPS \cite{usps} datasets.  Following recent literature on unsupervised learning \cite{ssfl}, we consider a smaller network of $N=10$ devices. We use a 4-layer convolutional encoder for the FMNIST dataset and a 3-layer fully connected encoder for the USPS dataset. For the unsupervised setting, we extend our method to allow at most two incoming edges as described in Remark \ref{remark:more_edges}. 

{For our multi-agent reinforcement learning setup, we train the policy for $5000$ iterations, using buffer size $\mathbf{\Omega}_i$ of $256$, global reward weight $\gamma = 0.5$, the reduction factor $\delta = 0.9$. We chose these parameters through our preliminary experimentation across tasks, considering their impact on the behavior of the algorithm. In particular, increasing the number of policy iterations allows the policy to benefit from more data, resulting in more beneficial choice of actions at the cost of increased D2D communication. The buffer size defines the history of samples that the policy learns from. A larger buffer size allows the policy to learn from older samples, generated by a more stale version of the policy.}

We express the probability of unsuccessful transmission $\bfP_D$ to $c_i$ from $c_j$  similar to \cite{semidecentralized} as \vspace{-2mm}
\begin{align}\label{eq:prob_drop}
     \bfP_D(i,j) = 1 - \exp \left(\frac{-(2^r - 1) \cdot \sigma^2}{\bfW_{i,j}} \right),
 \end{align}
where $r$ and $\sigma^2$ are the rate of transmission and noise power respectively and $\bfW \in \mathbb{R}^{N \times N}$, such that $\bfW_{i,j}$ defines the RSS \cite{fwc} at $c_i$ when it receives a signal from device $c_j$.



\textbf{Baselines:} We compare the performance of our algorithm with the following baselines. First, we consider (i) ``no exchange", which is suitable for applications where data exchange is prohibited or unsuitable, such as military or defense operations. Second, we consider (ii) ``closest", where graphs are generated when each receiver selects the transmitter with the highest probability of successful transmission, such as vehicle-to-vehicle (V2V) networks where distances between devices is dynamic. Third, we consider (iii) ``most trusted", where graphs are generated when each receiver selects the transmitter which can share the largest number of labels based on the inter-device trust matrix, such as on-body health monitors which communicate with paired smartphones. Fourth, we consider (iv) ``uniform", where graphs are generated using the Erd\H{o}s-Renyi model with uniform edge selection probability, such as a deployment of homogeneous wireless sensor nodes.
{Finally, we consider (v) ``GCN'', where a graph convolutional network (GCN) \cite{gcn} is trained at the server using the state and local data distributions as node features, and tasked with predicting directed edges between two devices. Additional information regarding the implementation of the GCN is provided in Appendix C. For fair comparison, each baseline is paired with the message passing algorithm (Alg. \ref{algo:message_passing})}.


\begin{figure*}[h!]	
        \begin{subfigure}[t]{0.48\textwidth}
		\centering
            \includegraphics[width=0.99\linewidth,height=3.5cm]{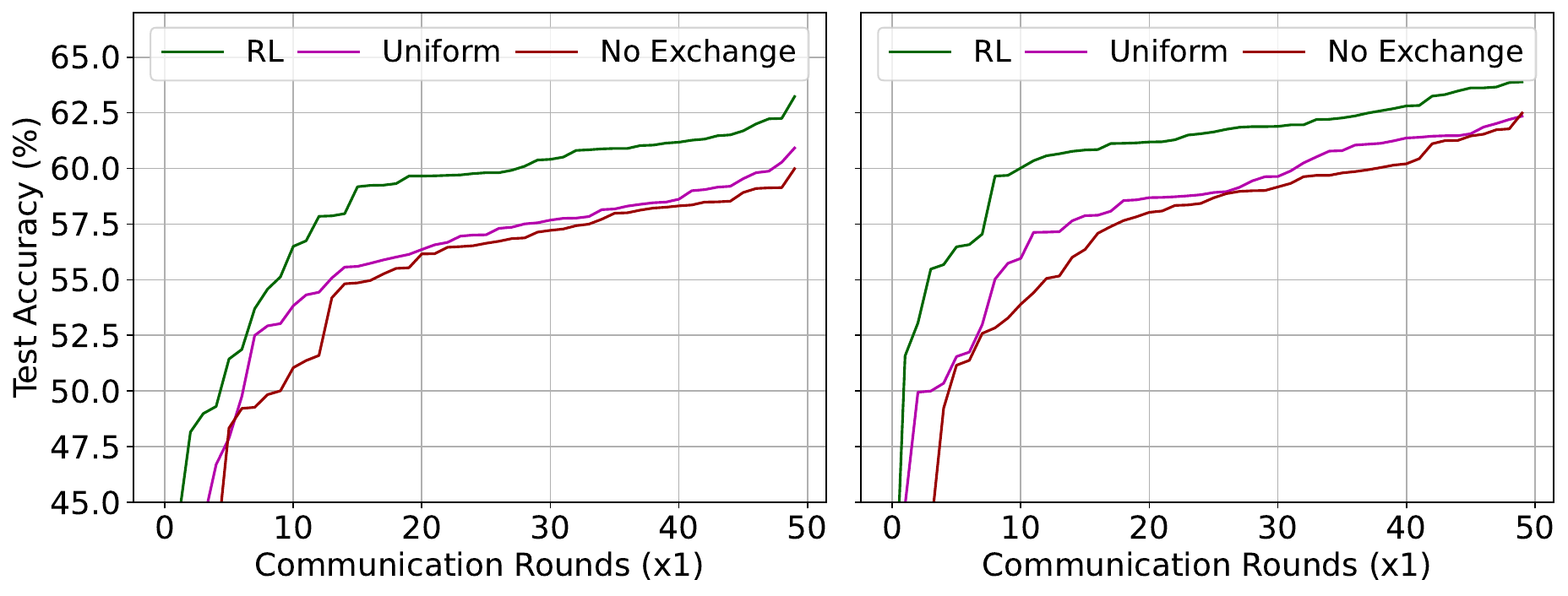}
		\caption{}\label{fig:7e}		
	\end{subfigure}
        \begin{subfigure}[t]{0.24\textwidth}
            \centering
            \includegraphics[width=0.99\linewidth,height=3.5cm]{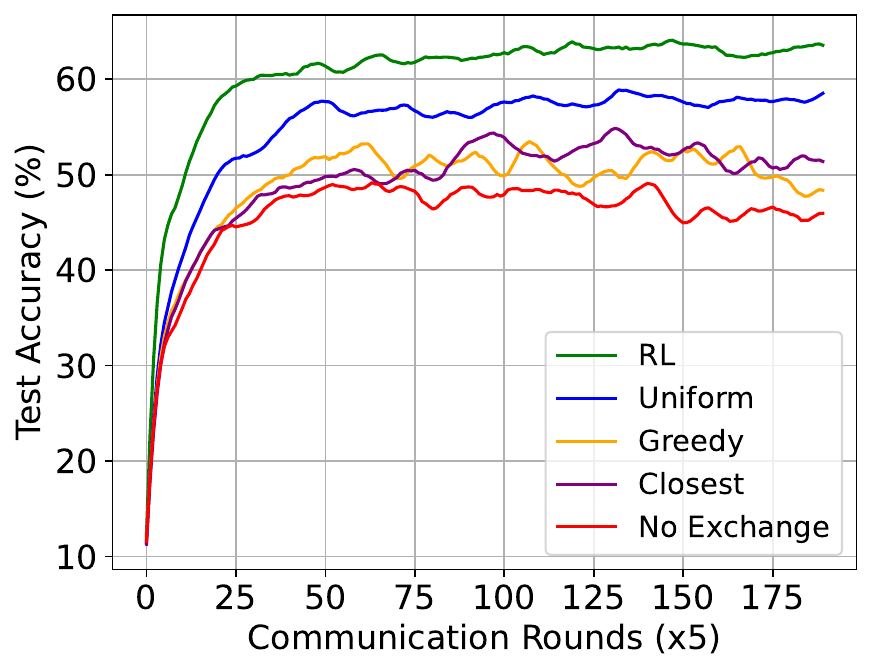}
            \caption{}\label{fig:r43_dec}
        \end{subfigure}
        \begin{subfigure}[t]{0.24\textwidth}
            \includegraphics[width=0.99\linewidth,height=3.5cm]{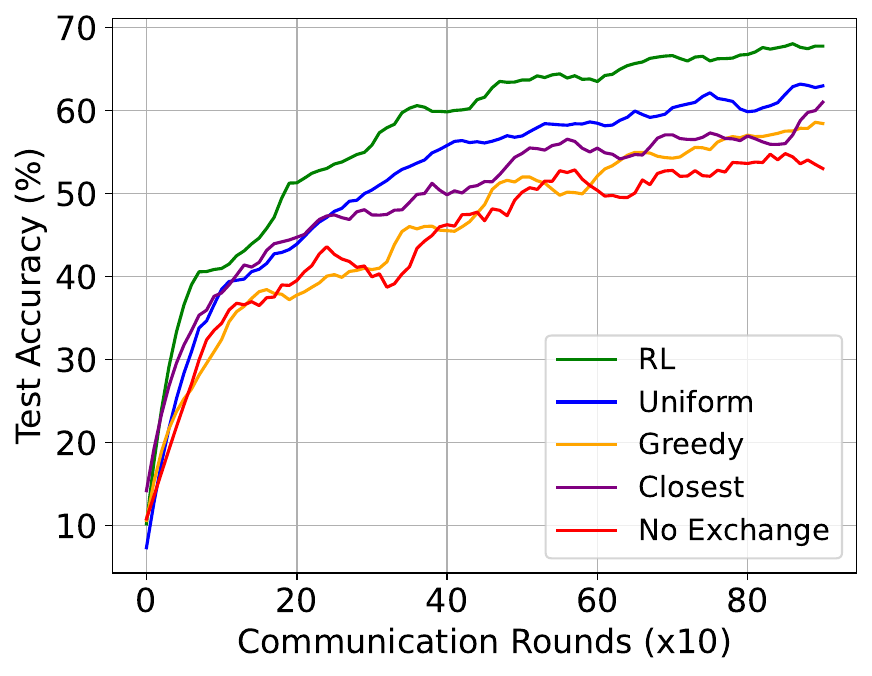}
            \caption{}\label{fig:r43_sdc}
        \end{subfigure}

        \centering
	\begin{subfigure}[t]{0.48\textwidth}
		\centering
		\includegraphics[width=0.99\linewidth,height=3.5cm]{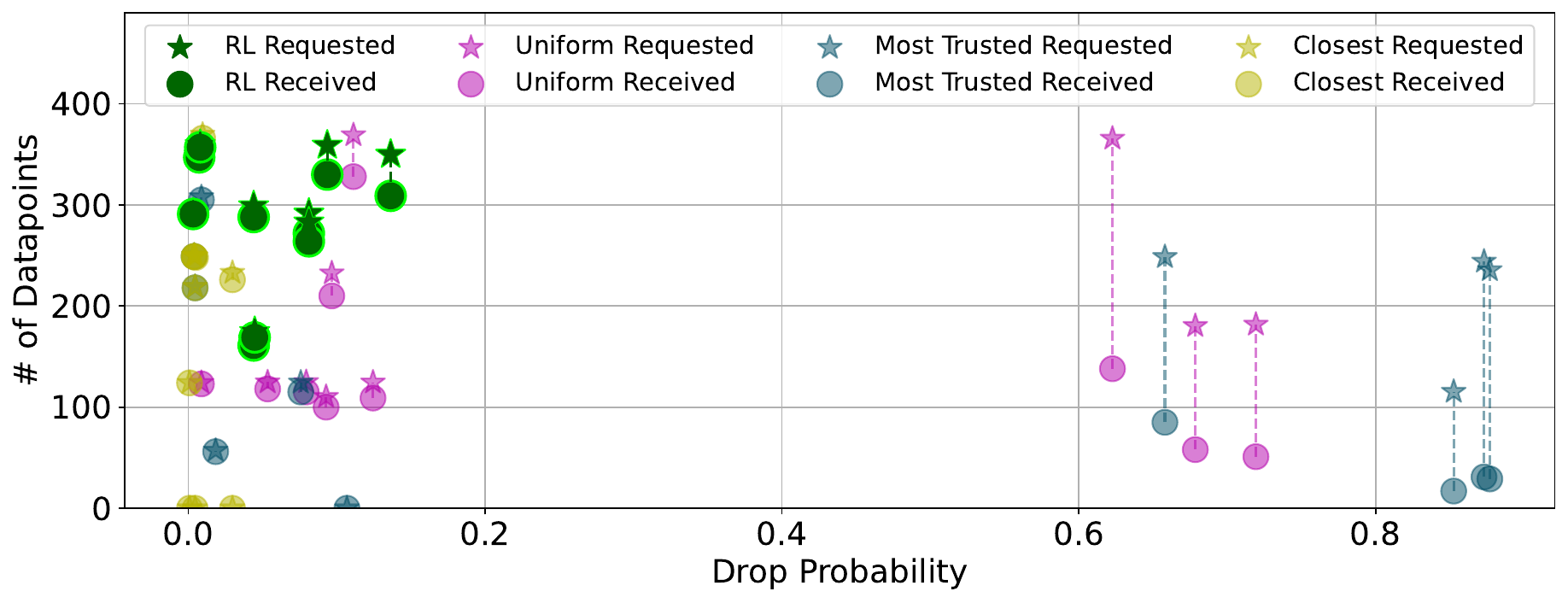}
		\caption{}\label{fig:7a}		
	\end{subfigure}
        \begin{subfigure}[t]{0.48\textwidth}
		\centering
            \includegraphics[width=0.99\linewidth,height=3.5cm]{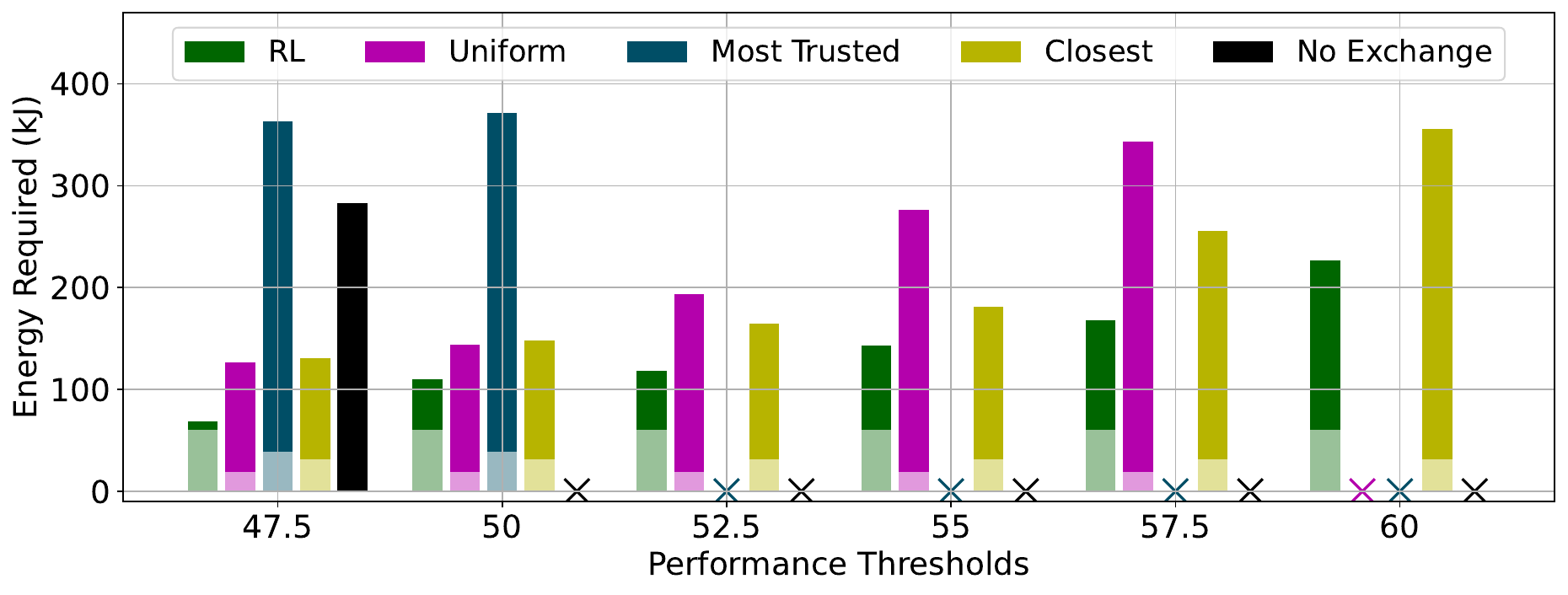} 
		\caption{}\label{fig:7b}		
	\end{subfigure}
\caption{In Fig. (a), we show that our method is compatible with popular federated aggregation algorithms such as FedProx (Left) and FedSGD (Right). In Fig. (b) and (c), we see that our method retains its performance advantages when executed as a pretraining step for fully decentralized and semi-decentralized downstream deployments, respectively. In Fig. (d), we see that our method significantly improves the overall probability of successful D2D transmission over baselines, while consuming less energy to reach performance milestones, which is shown in Fig. (e).}
 \vspace{-3mm}
 \label{fig:2}
\end{figure*}

\subsection{Qualitative Comparison of Graph Discovery Methods}
First, we investigate the ability of our method to discover D2D graphs which result in an optimal tradeoff between diversity improvement, trust and probability of successful transmission. We illustrate the impact of the choice of graph generation method on the produced graphs for the supervised paradigm in Fig. \ref{fig:gengraph}. The ``uniform" baseline randomly selects devices to connect to, and does not consider diversity improvements or probability of successful transmission, as seen by the large number of long-distance edges which are capable of sharing only a few labels. The ``closest" baseline selects devices with the highest probability of successful exchange of data, which is seen by the formation of highly-reliable short-distance edges. However, it does not consider the requirements of the receiving device or the trust between devices, resulting in minimal post exchange diversity improvement. The ``most trusted" baseline selects transmitters with the highest degree of trust, meaning that the transmitters can share datapoints from a variety of labels, as can be seen by the formation of exclusively high trust edges (green edges). However, in this case, the baseline does not consider the probability of successful transmission, dropping a large number of datapoints over unreliable long-distance transmitting edges as a result. In contrast to these baselines, our method discovers a graph which satisfies a tradeoff between the various system parameters which results in maximum improvements in local data diversity after D2D exchange, which can be seen by significant improvements in local diversity after data exchange (green nodes). 

\subsection{Performance on Different Datasets}
\label{sec:numexp_conv}
Now, we compare our algorithm on the CIFAR-10 dataset for the supervised case, RadioML for the semi-supervised case and the FMNIST dataset for the unsupervised case. {We also adapt our algorithm to regression tasks through a straightforward data partitioning step, and evaluate it using the California Housing dataset \cite{california_housing}.} For the unsupervised case, we use linear evaluation \cite{simclr} to obtain classification accuracy. We use the Federated Averaging \cite{fedavg} aggregation scheme for these experiments.

\subsubsection{Supervised Setting}
First, in \cref{fig:2a}, the plots illustrate that D2D information exchange using our method results in up to around $8\%$ improvement in the test accuracy over the competing baselines. Our approach finds a desirably structured D2D communication graph, resulting in considerable improvement of the FL performance over baselines. Our reward structure promotes improvement of local data diversity towards the ideal i.i.d., which subsequently accelerates the convergence of the global model due to increased alignment between local models. In contrast, baselines select links by heuristics which do not consider improvements in local diversity.

\begin{figure*}[h!]	
        \begin{subfigure}[t]{0.24\textwidth}
		\centering
            \includegraphics[width=0.99\linewidth,height=3.5cm]{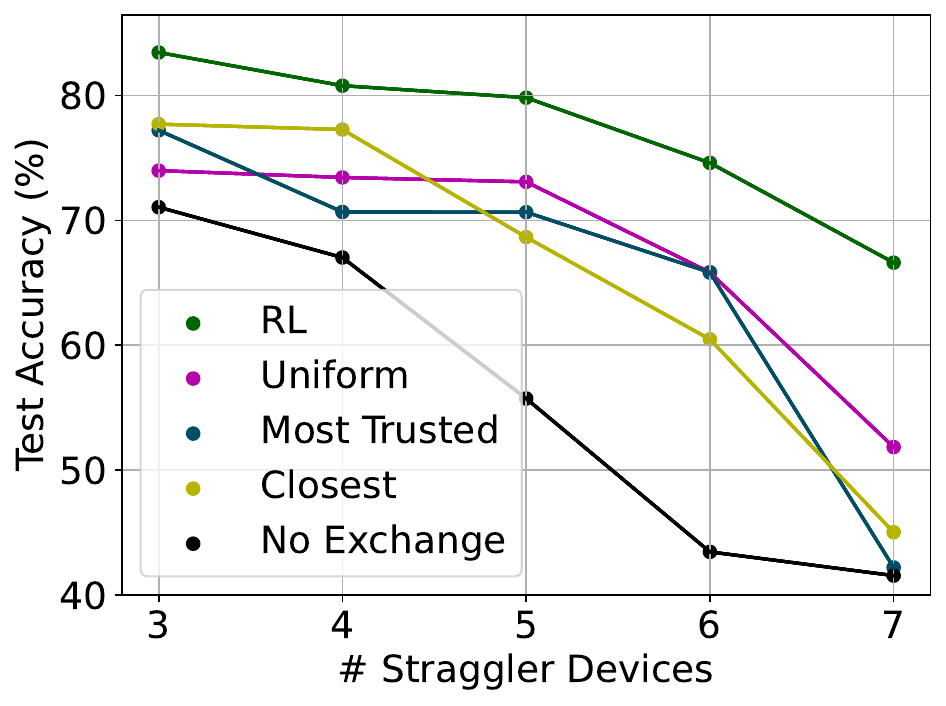} 
		\caption{}\label{fig:7c}		
	\end{subfigure}
        \begin{subfigure}[t]{0.24\textwidth}
		\centering
            \includegraphics[width=0.99\linewidth,height=3.5cm]{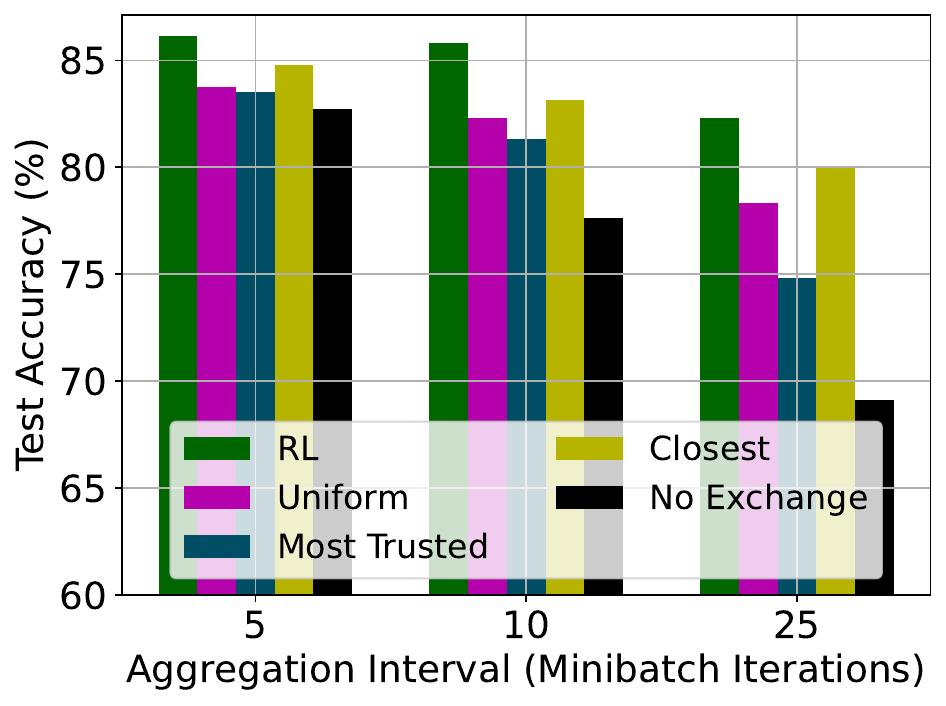}
		\caption{}\label{fig:7d}		
	\end{subfigure}
        \begin{subfigure}[t]{0.24\textwidth}
            \includegraphics[height=3.5cm,width=0.99\linewidth]{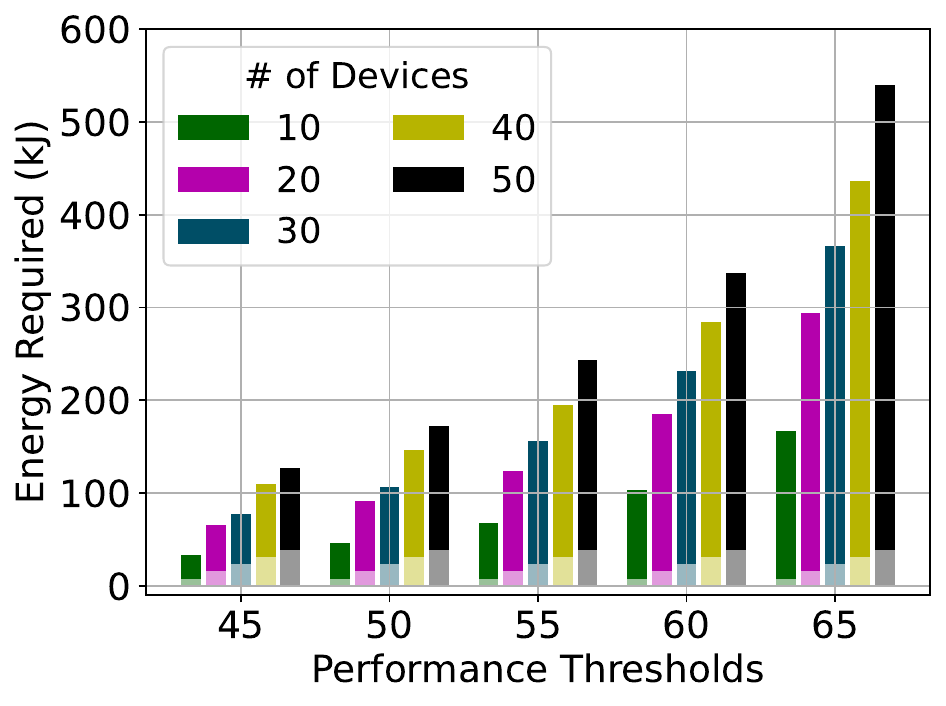}
            \caption{}\label{fig:f9_size}
        \end{subfigure}
        \begin{subfigure}[t]{0.24\textwidth}
        \includegraphics[height=3.5cm,width=0.99\linewidth]{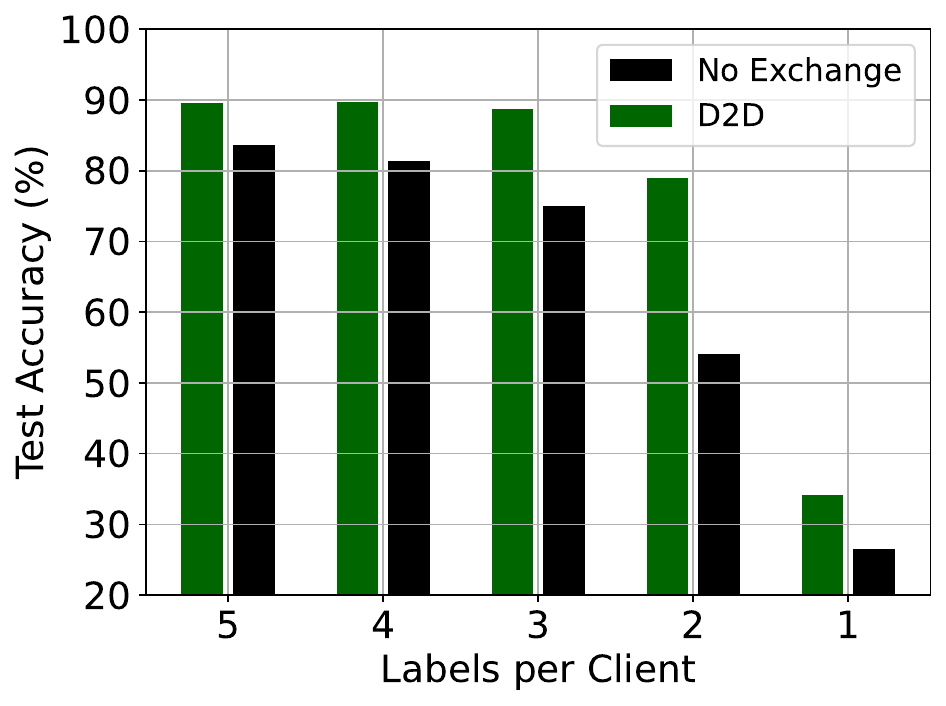}    
            \caption{}\label{fig:f9_skew}
        \end{subfigure}

\caption{In Fig. (a), we see that our method ensures that system performance is resilient to the presence of straggler devices. Performance also remains relatively consistent over larger aggregation intervals, as shown in (b). In Fig. (c), we see that the overall communication overhead of the system scales linearly with the number of devices for a given threshold of performance to be reached, thus retaining the advantages of our method over a broad range of system sizes. In Fig. (d), we see that for different levels of label skew, the relative performance improvements gained by our methods increase as the label skew at each device increases.}
 \vspace{-4mm}
 \label{fig:2}
\end{figure*}

\subsubsection{Semi-Supervised Setting}
Next, in \cref{fig:2f}, we observe that for the semi-supervised setting, our method results in improvement up to around $10\%$ in terms of test accuracy over the different baselines. This indicates that even for scenarios where data is sparsely labeled, our method improves local data diversity at each receiving device. Our method leverages the output of label propagation algorithms to decide links between devices based on updated label assignments using Alg. \ref{algo:message_passing}, such that local data diversity is improved. In contrast, baseline methods do not leverage the label assignments to inform link formations, resulting in worse performance.

\subsubsection{Unsupervised Setting}
Next, in \cref{fig:2d}, we observe that for the unsupervised setting, our method results in improvements up to $10 \%$ over baselines, indicating that for such scenarios, our method leads to better FL performance by improving the agreement between local data distributions and increasing diversity at each device. Our method performs local clustering in a globally consistent subspace, enabling meaningful comparisons between local and remote clusters. We leverage this to enable transmitters to identify clusters crucial to improve data diversity at the receiver, as opposed to baselines which do not use any measure of importance.

{\subsubsection{Extension to Regression Setting}
Finally, in \cref{fig:2h}, we consider how the performance of our method translates to a regression task. We consider the California Housing dataset \cite{california_housing}, and first choose the number of partitions $L$ to split the local data at each device into $L$ clusters. This is done by sorting all local datapoints in ascending order of their labels, and then finding the top $L-1$ largest differences between two consecutive labels. We observe that our method converges after $\sim 25$ communication rounds, while baselines require at least $\sim 35$ rounds to converge. This illustrates that, with an appropriately defined data partitioning, our method retains performance advantages when extended to regression tasks.} 

For additional experimental results on RadioML, CIFAR-10, SVHN and USPS datasets, see Appendix D.

\subsection{Performance on varying FL Schemes}
Next, we apply our method to two other popular FL schemes: FedProx \cite{fedprox} and FedSGD \cite{fedavg} for the unsupervised learning case on the Fashion-MNIST dataset for $10$ devices, and compare the results in the left and right plots respectively of Fig. \ref{fig:7e}.
We observe that our method reaches $60 \%$ testing accuracy $2 \times$ to $3 \times$ faster than baselines with an overall improvement of around $4 \%$, which indicates that it can be applied over different FL schemes without sacrificing performance gains. In FedProx, our method results in more consistent proximal regularization due to a more diverse set of local data post-exchange, improving model performance. In FedSGD, our method results in more i.i.d local data  post-exchange, resulting in more homogenous model gradients, accelerating the convergence of the global model as is seen in Fig. \ref{fig:7e}. This shows that our framework can be adapted to different FL methods.

{\subsection{Performance on Decentralized Downstream FL Schemes}
Next, we use our method as a pretraining process for downstream fully decentralized \cite{decentralized_r43} and semi-decentralized \cite{sdc_r43} FL schemes. We consider the supervised CIFAR-10 dataset with $25$ devices in Fig.~\ref{fig:r43_dec} and Fig. \ref{fig:r43_sdc} respectively. In the fully decentralized learning case, each device aggregates local models from $7$ neighbors at random. For semi-decentralized learning, we consider a scheme similar to \cite{sdc_r43}, using disjoint subsets of $5$ devices. Local models are exchanged only between devices belonging to the same subset after every $2$ minibatch iterations. We perform global aggregations every $8$ minibatch iterations, where the server aggregates models from $1$ device chosen uniformly at random from each subset, and broadcasts the aggregated model to all devices. We observe that, for both the fully decentralized and semi-decentralized settings, our method outperforms all associated baselines by at least $\sim 5\%$. This demonstrates the ability of our method to improve downstream FL tasks in the absence of a server, as all message passing, reward calculation and policy prediction steps can be done in a decentralized manner. For the semi-supervised learning paradigm, this illustrates the ability of our method to identify smaller subgraphs for disjoint sets of devices.}

\subsection{Reliability of D2D Performance} Next, we study D2D reliability in terms of the probability of successful transmission. The corresponding results are shown in Fig. \ref{fig:7a}. We consider the semi-supervised setting using the RadioML dataset with $10$ devices. We observe that our method consistently predicts links to reduce inter cluster communication while improving system performance, while baseline methods either request a large number of datapoints from unreliable links (``most trusted") or select strong links between devices which can only share a limited amount of information (``closest"). In practice, this results in reduced communication overhead compared to baselines, thus saving additional costs required to ensure successful transmission over unreliable channels.

\begin{figure*}[h!]
    \begin{subfigure}[t]{0.24\textwidth}
        \centering
        \includegraphics[width=0.99\linewidth,height=3.5cm]{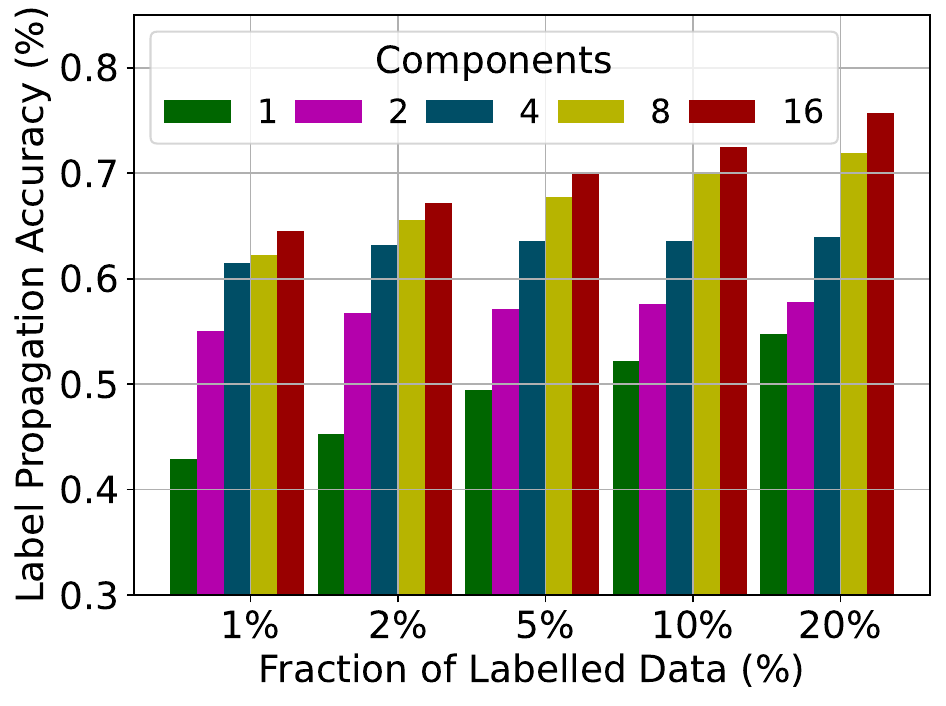}
    \caption{}\label{fig:7g}		
    \end{subfigure}
    \begin{subfigure}[t]{0.24\textwidth}
        \centering
        \includegraphics[width=0.98\linewidth,height=3.5cm]{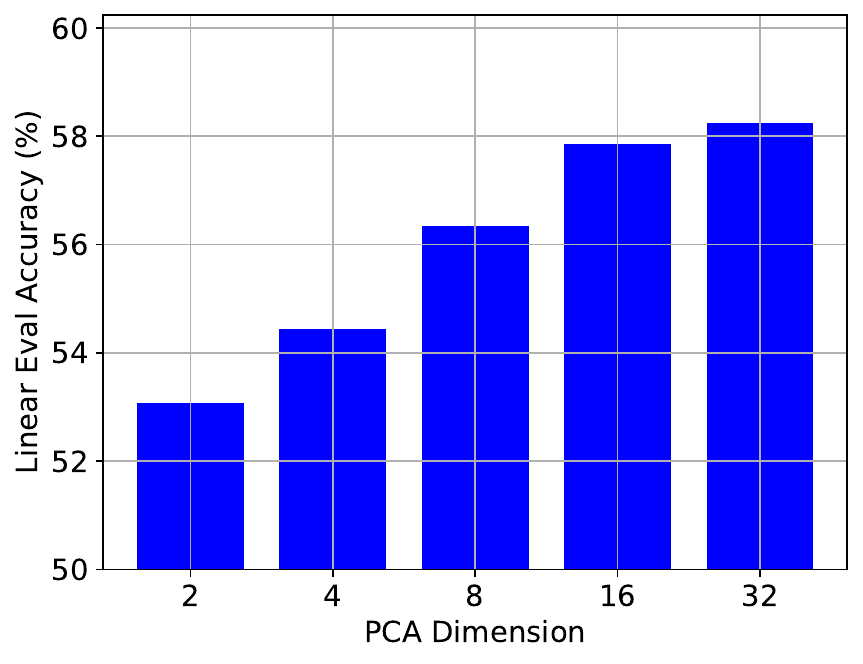}
        \caption{}\label{fig:f8_pca}
    \end{subfigure}
    \begin{subfigure}[t]{0.24\textwidth}
        \includegraphics[width=0.98\linewidth,height=3.5cm]{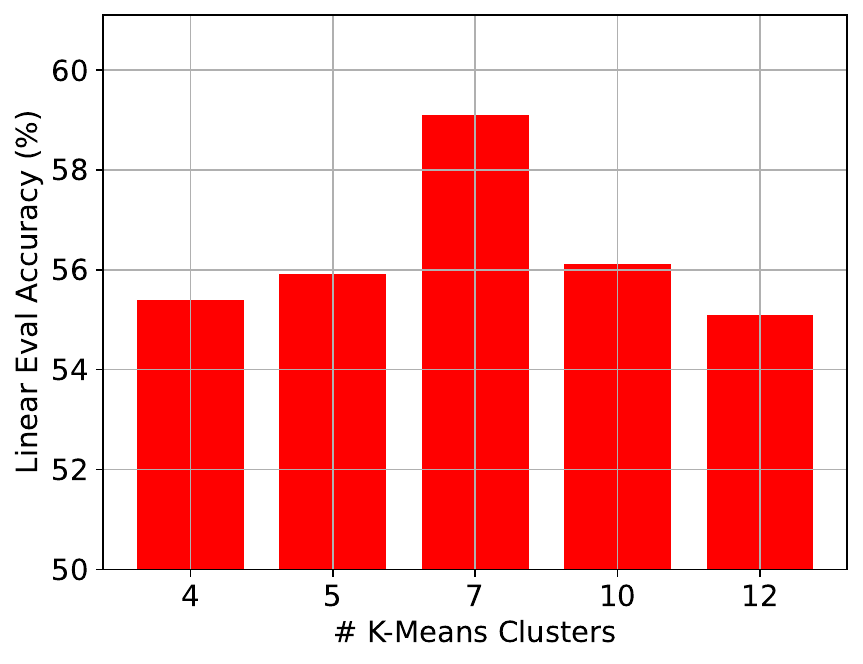}
        \caption{}\label{fig:f8_kmeans}
    \end{subfigure}
    \begin{subfigure}[t]{0.24\textwidth}
    \centering
        \includegraphics[width=0.99\linewidth,height=3.5cm]{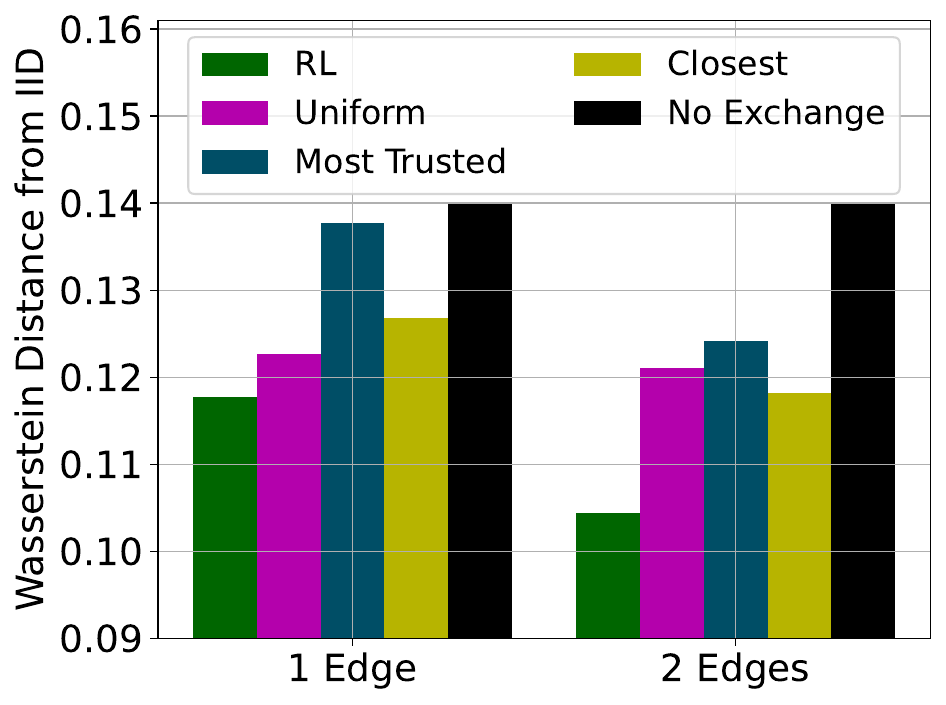}
    \caption{}\label{fig:7f}		
    \end{subfigure}
\label{fig:f8}
\vspace{-0.05 in}
\caption{For the semi-supervised case, in Fig. (a) the number of PCA components used for label propagation defines a tradeoff in terms of communication overhead and labeling accuracy. For the unsupervised case, in Fig. (b), we see diminishing returns in the performance gained by increasing the dimension of PCA components, while in Fig. (c), we observe a clear optimum value for the number of K-Means clusters to be used such that each cluster is homogenous as well as large enough to facilitate data exchange. In Fig. (d), our method chooses data to exchange such that post-exchange distributions are significantly closer to the ideal i.i.d scenario, even for multiple incoming edges.}
\vspace{-0.1 in}
\end{figure*}


\subsection{Energy Consumption to Reach Benchmarks} Next, we compare the energy required by our method to achieve performance benchmarks with baselines. We consider the semi-supervised setting using the RadioML dataset with $25$ devices. We use the wireless energy consumption model in \cite{eed2d} to calculate the energy consumed for D2D and device-to-server (D2S) communication. In this simulation, we assume that the D2S distance is $3\times$ the average D2D distance. Fig. \ref{fig:7b} shows that our method uses up to $\sim 5 \times$ less energy to reach benchmarks as the baselines, despite the initial overhead due to D2D exchange. We observe that our method uses the same amount of energy to achieve performance improvements of $\sim 5\%$ over the ``closest" and ``uniform" baselines and more than $10 \%$ over the other baselines. Note that suboptimal links cause fewer datapoints to be exchanged for baselines, resulting in lower D2D energy, but significantly higher D2S energy.

\subsection{Effect of Stragglers on Performance} We now study the performance of our method in the presence of straggler devices \cite{wang2024device} in the FL system, which do not participate in model aggregation. As the number of stragglers increases, fewer local models are aggregated. As each model is biased towards non-i.i.d local data, it reduces the accuracy of the global model. We consider the supervised setting using the SVHN dataset with $15$ devices.
\label{sec:strg_res}
In Fig. \ref{fig:7c}, we choose stragglers randomly and show that our method is more resilient than the baselines by incurring a performance penalty of only $\sim 14 \%$ compared to the baselines whose performance deteriorates by $\geq 20 \%$. It indicates the ability of our method to share data that makes up for the bias in the aggregated model as a result of stragglers, making it inherently robust to node failure and heterogeneous communication capabilities. 

\subsection{Change in Aggregation Interval}Next, we observe the effect of various aggregation intervals $\tau_a$, or the frequency of model synchronization. A low $\tau_a$ can result in faster convergence, but involves a larger overhead due to more frequent D2S communication. We consider the supervised setting using the SVHN dataset. Fig. \ref{fig:7d} shows that our method outperforms the baselines by a considerable margin when $\tau_a$ becomes larger, which indicates its resilience to delays in model aggregation and a lower local model drift. Thus, a small initial overhead for our method results in significant reduction in D2S overhead by retaining similar performance. 
\vspace{-2mm}
{\subsection{Effect of Variation in System Size} In Fig.~\ref{fig:f9_size}, we analyze the effect of the number of devices in the system on the D2D energy overhead introduced by our method. We conduct experiments for the supervised CIFAR-10 dataset and observe the energy required for D2D communication (light bottom bar) and D2S communication (dark top bar) to reach different performance thresholds for a varying number of devices (see Fig.~\ref{fig:energy_svhn} in Appendix D for results in SVHN). We observe that for both datasets, the overall communication overhead of the system scales linearly with the number of devices for a given threshold of performance to be reached, thus retaining the advantages of our method over a wide range of system sizes. This also matches the expected algorithmic complexity. Specifically, for each iteration of policy training, $N$ devices choose exactly $1$ incoming edge, and the policy training process is executed sequentially for $E$ edges. Thus, the total D2D energy consumption during RL training is proportional to $O(NEt)$, where $N$ is the number of devices, $E$ is the maximum number of incoming edges at each device and $t$ is the number of RL iterations for which the policy is trained. Thus, when $E$ and $t$ are constant, the D2D energy consumption increases linearly with $N$.}     

\vspace{-2mm}
{\subsection{Effect of Variation in Initial Label Skew} In Fig.~\ref{fig:f9_skew}, we show the effect of label skew on the downstream federated learning performance for the supervised SVHN dataset with $10$ devices. We vary the number of labels initially available at each device and compare two scenarios, (i) where no data is exchanged and (ii) where data is exchanged using our method. We observe that the increments in performance of our method become more significant as the number of labels at each device decreases (i.e, the label skew increases). The improvement is most significant for settings with the highest label skew, as seen by the $\sim 1.4 \times$ gain in test accuracy when each device has data from $1$ label before D2D exchange, compared to the $\sim 1.1 \times$ gain when each device has data from $5$ labels before D2D exchange. In settings with a significant degree of label skew, D2D exchange can improve data diversity significantly. We present additional results for CIFAR-10 in Appendix D (Fig. 15).} 
\vspace{-2mm}
\subsection{Performance of Distributed Label Propagation} In Fig. \ref{fig:7g}, we observe the accuracy of the distributed label propagation algorithm for labeling tasks in sparsely labeled datasets. We consider the semi-supervised RadioML dataset with $15$ devices. We vary the number of distributed PCA components being shared between devices, which reflect the data bandwidth utilized as well as the degree of data-specific information being exposed to other devices. We also vary the fraction of unlabeled data to observe the performance of the algorithm in extreme cases. 
We observe that the labeling accuracy increases with the number of components exchanged, which characterizes a tradeoff in terms of communication overhead or local information exposure and prediction accuracy.

{\subsection{Changing the Dimensionality of PCA} In Fig.~\ref{fig:f8_pca}, we show the effect of changing the dimensionality of the PCA components on the linear evaluation accuracy for the FashionMNIST dataset in the unsupervised setting. We observe that the performance gain observed by increasing the number of PCA components diminishes as the dimensionality is increased. This is consistent with the PCA components being chosen in rank-order based on the magnitude of the associated singular values, which indicates the variance explained by the chosen principal components; the additional variance in the data explained by subsequent PCA components diminishes as more PCA components are added.

\subsection{Changing the Number of K-Means Clusters} In Fig.~\ref{fig:f8_kmeans}, we show the effect of changing the number of K-Means clusters for the FashionMNIST dataset in the unsupervised setting. We observe that the system performs best in this setting with $7$ clusters, diminishing above and below this point. With too few clusters, the data contained in each cluster is not homogeneous, and the distribution of data within the cluster cannot be accurately described by the shared parameters. On the other hand, if the number of clusters becomes too large, each cluster only contains a few datapoints, which prevents the transmitter from sharing data to maintain the threshold conditions described in \eqref{eq:data_diversity}.} 

\begin{figure}
    \centering
    \begin{subfigure}[t]{0.24\textwidth}
        \includegraphics[height=3.15cm]{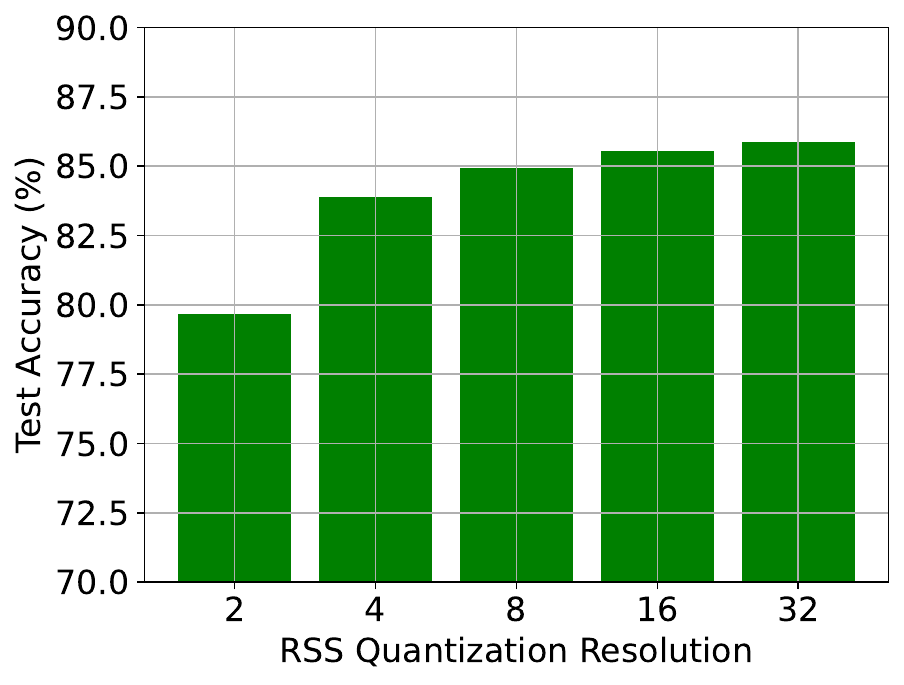}
        \caption{}\label{fig:a_rss}
    \end{subfigure}
    \begin{subfigure}[t]{0.24\textwidth}
        \includegraphics[height=3.15cm]{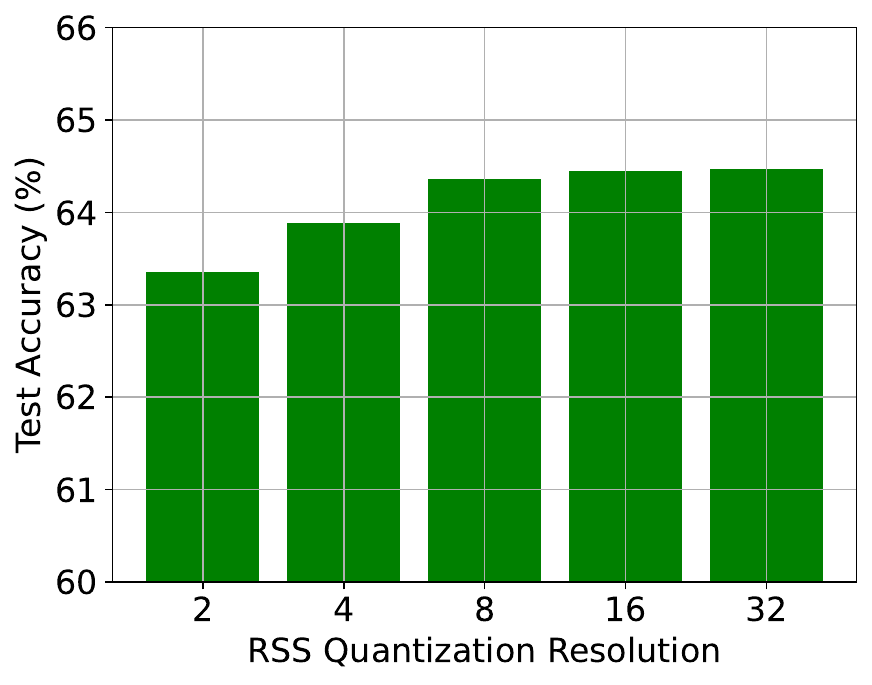}
        \caption{}\label{fig:a_rss_cifar}
    \end{subfigure}
    \vspace{-0.07 in}
    \caption{For dynamic wireless scenarios, the FL performance improves as the quantization resolution is increased, as policies are able to distinguish between RSS values to higher levels of granularity.}
        \vspace{-0.1 in}
    \label{fig:rss}
\vspace{-2mm}
\end{figure}

\vspace{-0.06 in}
\subsection{Effect of Multiple Edges}

Next, we observe the effect of predicting additional edges on the local data bias after D2D exchange in Fig. \ref{fig:7f} by comparing the average 1-Wasserstein distance between the ideal i.i.d distribution and the local data at each device after D2D exchange for the unsupervised FMNIST dataset. The label information is only used for distance calculation.  

We observe that our method produces local distributions which are closest to the ideal i.i.d compared to baselines, indicating a larger reduction in local biases even in the absence of label information. Further reduction in bias by adding an edge provides a diminishing return, as our method achieves reductions of around $17\%$ followed by an additional reduction of around $5\%$ after the formation of the first and second edges, respectively. 
This indicates that our method maximizes the reduction in local biases through the formation of a single edge by exchanging information crucial to diversity maximization. 

\vspace{-0.06 in}
{\subsection{Performance in Dynamic Wireless Scenarios} Next, we illustrate the ability of our method to adapt to dynamic drop probabilities $\mathbf{P}_D(i,j)$ for the SVHN and CIFAR-10 datasets in Fig.~\ref{fig:a_rss} and Fig.~\ref{fig:a_rss_cifar} respectively. We consider $25$ devices with $3$ labels each, and allow $\mathbf{P}_D$ values to change through the course of training, by varying the the RSS $\{\textbf{W}_{i,j}\}_{i,j \in [0,N]}$ at every step of the RL training process. We vary $\textbf{W}_{i,j}$ by sampling from the $\mathcal{N}(0.3,0.1)$ Gaussian distribution, and assume $r=0.8$ and $\sigma^2=0.02$ in (\ref{eq:prob_drop}). We then quantize $\textbf{W}_{i,j}$ using the given resolution over the range of the Gaussian distribution, truncating it between $(0.05,0.55)$, and observe the effect of using different levels of quantization to discretize the RSS. This sampling and quantization process is performed at every RL training step for each device $c_i$, producing $\bfs_i^q = \{\bfW_{i,j}:c_j \in \mathcal{C}\}$, which is an instance of the $q$-th unique state at device $c_i$, as defined in Sec. IV-B of the main manuscript. We observe that the downstream FL performance improves with an increase in the quantization resolution. The larger number of states allow the policies to distinguish between RSS values to a higher level of granularity, resulting in more accurate policy predictions.} 

\begin{figure}
    \centering
        \includegraphics[width=0.99\linewidth,height=3.5cm]{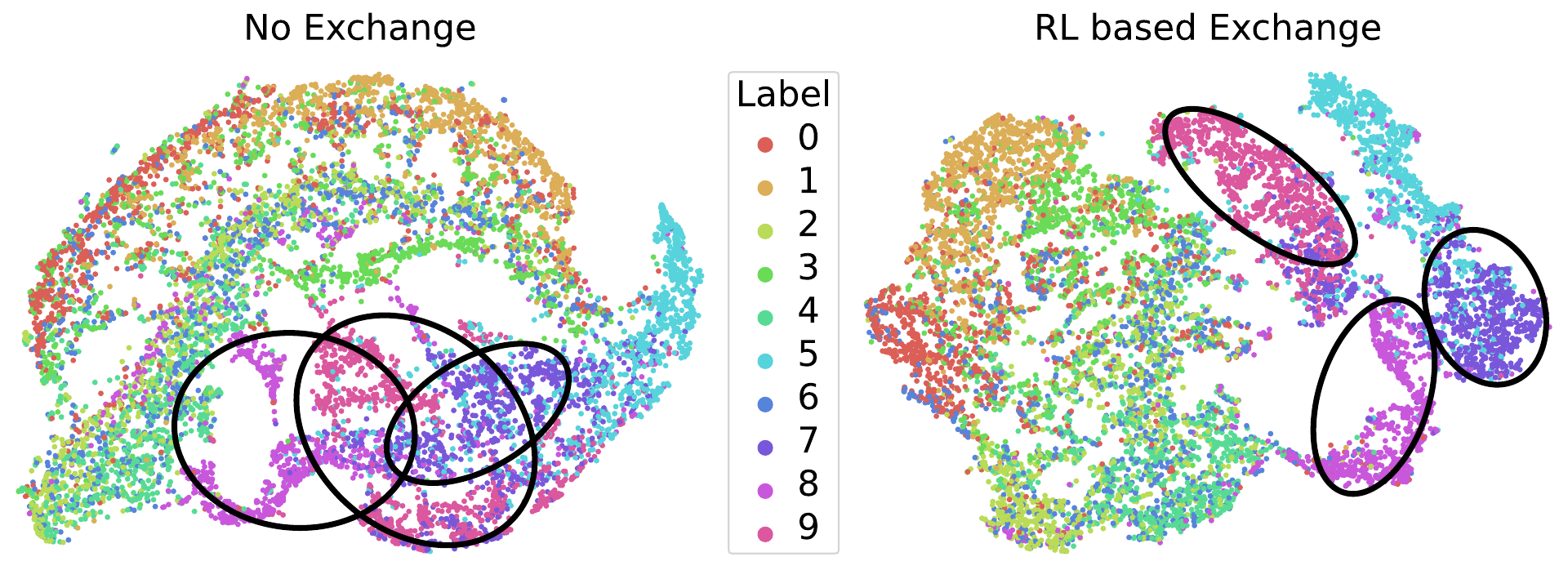}
        \label{fig:7h}		
    \caption{Unsupervised D2D exchange using our method results in well separated clusters, improving downstream classification performance.}
    \label{fig:7h}		
\vspace{-4mm}
\end{figure}
\subsection{Separation in Embedding Spaces in Unsupervised Scenarios}

In Fig. \ref{fig:7h}, we plot the embeddings generated by the global model in 2-D space using t-SNE \cite{tsne}. We consider the unsupervised learning scenario for the FMNIST dataset. We observe that our method promotes tightly clustered global embeddings, enabling easier downstream classification, thereby improving the performance of global models. We observe that our method creates well separated clusters of labels $5,7$ and $9$, corresponding to Sneakers, Sandals and Ankle Boots in the dataset, which are visually similar and thus hard to distinguish. When no data is exchanged, however, these clusters overlap significantly, increasing the chance of incorrect prediction. 

\section{Conclusion}
In this paper, we developed a novel framework for inter-device cooperation in D2D enabled FL to improve local data diversity while being cognizant of inter-device trust rules and communication efficiency. 
We utilized decentralized multi-agent RL to train independent policies at each device, which collaboratively learn an optimal D2D communication graph over the system. 
We designed reward functions specific to the multiple learning paradigms and a lightweight message passing system to facilitate policy training without significant communication overhead or exposure of local data to the server. 
We empirically showed that our method discovers D2D graphs which significantly improve performance on popular datasets in terms of accuracy and energy efficiency.
Our work can be extended to the concurrent (i.e., non-greedy) discovery of multiple incoming edges, whose impact will depend on the degree of system heterogeneity. 
Also, as our framework is a pre-training procedure, our metrics consider the importance of data independent of model parameters. Online cooperation between devices will differ significantly in its implementation. 

\bibliographystyle{IEEEtran}
\bibliography{gbcm_v2.bib} 

\newpage

\begin{appendices}

\section{Motivating Example for Message Passing}
\label{appx:mot_ex}

\begin{figure}[h!]
    \centering
    \includegraphics[width=0.95\columnwidth]{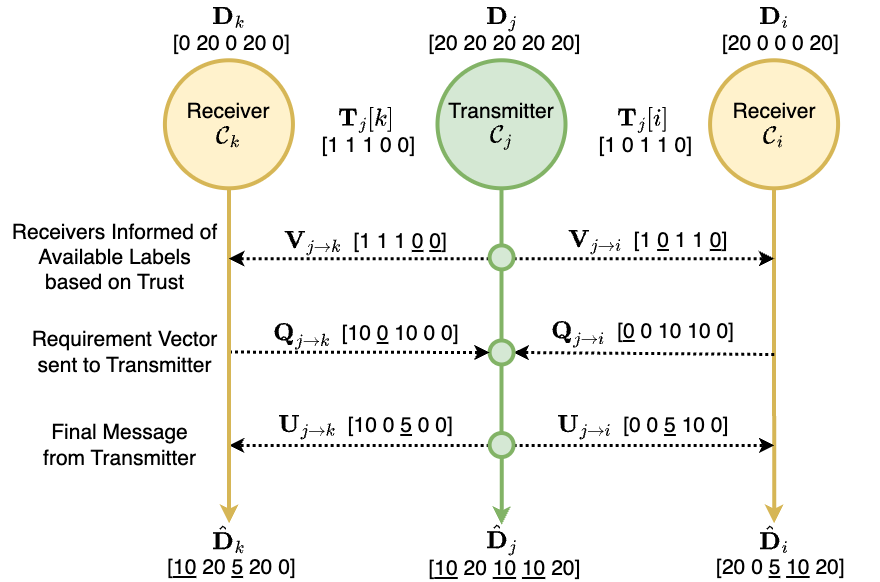}
    \caption{An example of message passing process for $\bfb_n[\ell] = 10 ~\forall~\ell,n$.}
    \label{fig:message_passing}
    \vspace{-4mm}
\end{figure}

\begin{example} 
{\normalfont 
We consider an example with $L = 5$ classes in Fig. \ref{fig:message_passing} to clarify the message passing algorithm. Assume that $c_i$ and $c_k$ have their corresponding class-distribution vector $\bfD_i = [20 \; 0 \; 0 \; 0 \; 20]$ and $\bfD_k = [0 \; 20 \; 0 \; 20 \; 0]$. Next, according to the link formation procedure, we assume that device $c_j$, with $\bfD_j = [20 \; 20 \; 20 \; 20 \; 20]$ is supposed to transmit datapoints to both $c_i$ and $c_k$. Moreover, the corresponding trust vectors for $c_i$ and $c_k$ are $\bfT_j[i,:] = [1 \; 0 \; 1 \; 1 \; 0]$ and $\bfT_j[k,:] = [1 \; 1 \; 1 \; 0 \; 0]$. Now if we assume $\bfb_i = \bfb_j = \bfb_k = [10 \; ... \; 10]$, after the datapoint exchange following our message passing algorithm, the updated class distribution vector will be, $\hat{\bfD}_i = [20 \; 0 \; 5 \; 10 \; 20]$, $\hat{\bfD}_j = [10 \; 20 \; 10 \; 10 \; 20]$ and $\hat{\bfD}_k = [10 \; 20 \; 5 \; 20 \; 0]$. A detailed illustration is provided in  Fig. \ref{fig:message_passing}.

Note that device $c_j$ (i) shares datapoints only from trusted classes with $c_i$ and $c_k$, and (ii) retains enough for its own threshold constraints to be satisfied as well. 
Consider $\ell = 4$, where $\bfT_j[i,\ell]=1$ and $\bfT_j[k,\ell]=0$. Thus, device $c_j$ conveys to $c_i$ that it can share data from label $\ell=4$ through $\bfV_{j \rightarrow i}$ and to $c_k$ that it cannot share it through $\bfV_{j \rightarrow k}$. 
Now, consider $\ell=3$, where the total demand $\bfQ_{j \rightarrow k} + \bfQ_{j \rightarrow i} = 20$, is greater than what is available at $c_i$ to share, which is $\bfV_{j \rightarrow i}[\ell] = \bfV_{j \rightarrow k}[\ell] = 10$. Hence, the demand is split as $\bfU_{j \rightarrow k}[\ell] = \bfU_{j \rightarrow i}[\ell] = 5$, leaving $c_j$ with enough datapoints $\hat{\bfD}_{j}[\ell] = \bfb_j [\ell] = 10$.}
\end{example}

\begin{remark}
{\normalfont
Note that in the supervised and semi-supervised case, the data distribution $\bfD_j$ of any transmitter $c_j$ is never fully exposed to a receiver $c_i$, unless $\bfT_j[i,k] = 1 ~\forall ~ k$ (complete trust). Also, due to (\ref{eq:tx_buffer}), $c_j$ may want to share fewer datapoints from a class $\ell$ with requesting devices, as $c_j$ must be left with at least $\bfb_j[\ell]$ datapoints after each exchange.
}
\end{remark}

\section{Motivating Example for the Diversity Metric}
\label{appx:mot_ex_unsupervised}

\begin{example} 
{\normalfont 
As a motivating example, we calculate the system agreement score between pre-exchange and post-exchange data distributions with varying \emph{change} in the degree of non-i.i.d after data exchange. For the purposes of analysis, we assume that the ground truth labels are observable (\emph{we do not use the ground truth labels in our reward calculations}). We measure the change in degree of non-i.i.d using the concentration parameter $\alpha$ of the Dirichlet distribution over the labels at each local dataset before and after data exchange (as a distribution becomes more i.i.d, $\alpha$ increases). We now evaluate the system agreement reward for scenarios with varying changes in $\alpha$ before and after data exchange. The results are shown in Table \ref{tab:dirich_comp}, and we observe that the system agreement score increases as the final distribution becomes more i.i.d, and produces a negative reward if it becomes more non-i.i.d.
}
\end{example}

\begin{table}[t!]
\centering
\small
\begin{tabular}{|l l|l|l|l|l|}

\hline
& & \multicolumn{4}{c|}{Post-Exch. Dist} \\ \cline{3-6}
 & & $\alpha = $ 0.01 & $\alpha = $ 0.1 & $\alpha = $ 1.0 & $\alpha = $ 10.0\\
\hline

 &$\alpha = $ 0.01 & -0.0369 & 0.100 & 1.906 & 4.515 \\ 
Pre- &$\alpha = $ 0.1 &  -0.3019 & -0.1008 & 1.016 & 3.984\\
Exch. &$\alpha = $ 1.0 &  -0.9432 & -0.8208 & 0.2300 & 3.452\\ 
Dist. &$\alpha = $ 10.0 & -0.9999 & -0.9999 & -0.9736 & 0.385\\ 
\hline
\end{tabular}
\caption{We compare the system agreement reward between pre-exchange and post-exchange distributions with varying degrees of non-i.i.d-ness defined by the Dirichlet parameter $\alpha$. Our system agreement formulation promotes data exchange such that post-data exchange distributions are more i.i.d. than pre-data exchange distributions. }
\label{tab:dirich_comp}
\end{table}

\begin{remark}
{\normalfont
Intuitively, the improved diversity of the post-exchange distributions, characterized by $\hat{\bfD_i}$ for supervised and semi-supervised cases and $\sum_{k=1}^L \frac{\Tr{(\tilde{\Sigma_i^k}})}{\Tr(\Sigma_i^k)}$ for the unsupervised case, mitigates the detrimental effect of straggler devices \cite{wang2024device} in the system by ensuring that datapoints of any class are present at more devices. We elaborate on this effect in Sec. \ref{sec:strg_res}.
}
\end{remark}

\section{Details on Simulation Setup}
\label{appx:sim_details}

For the supervised and semi-supervised federated learning scenarios, we use the RadioML \cite{radioml}, CIFAR-10 and SVHN  datasets with a $80/20$ split to obtain training and testing datasets respectively. We consider a network of $N = 25$ devices, and emulate non i.i.d training data across all devices. Each device has $990$ samples for RadioML, and $1200$ for the CIFAR-10 and SVHN from $4$ different classes. We use the Alexnet \cite{alexnet} architecture as the FL model for CIFAR-10, RadioML and SVHN datasets. For the semi-supervised setting, we assume that $15\%$ of the data is labeled. Note that we allow at most one incoming edge for the supervised and the semi-supervised setting. 

For the unsupervised federated learning scenario, we use the Fashion-MNIST and USPS \cite{usps} datasets with a $80/20$ split identical to the supervised case. We augment the USPS dataset with left and right horizontal rotation views and resized crops in order to make the contrastive learning problem more challenging to solve. We consider a network of $N=10$ devices, which is a reasonable scenario for unsupervised learning suggested by recent literature \cite{ssfl}. Each device has $6000$ samples for FMNIST and $2600$ samples for USPS. We use a 4-layer convolutional encoder for the FMNIST dataset and a 3-layer fully connected encoder for the USPS dataset. Both encoders have an embedding dimension of $8$. Note that in the unsupervised setting, we extend our graph discovery method to multiple edges, as mentioned in Remark \ref{remark:more_edges}.

{We assume that each device has data from $3$ labels, chosen randomly for each device. Each device is allocated datapoints such that their local datasets are comprised of $70\%$, $20\%$, and $10\%$ from each of the $3$ labels. For a given number of devices, the complete dataset is divided such that each device has approximately the same number of datapoints.}

We consider a network architecture similar to \cite{semidecentralized}, which assumes D2D communication conducted using OFDMA. We assume similar noise power $\sigma^2$ across all channels and a constant rate of transmission $r$ between devices. Thus, we express the probability of unsuccessful transmission $\bfP_D$ to $c_i$ from $c_j$  similar to \cite{semidecentralized} as \vspace{-2mm}
\begin{align}\label{eq:prob_drop_appdx}
     \bfP_D(i,j) = 1 - \exp \left(\frac{-(2^r - 1) \cdot \sigma^2}{\bfW_{i,j}} \right),
 \end{align}
where $\bfW \in \mathbb{R}^{N \times N}$, such that $\bfW_{i,j}$ defines the RSS at $c_i$ when it receives a signal from device $c_j$. As described in Sec. \ref{sec:system_model}, a high probability of unsuccessful transmission results in a large number of datapoints dropped during D2D exchange, which negatively affects the performance of the system. {It should be noted that a change in the physical layer characteristics (e.g., fading type, modulation scheme, coding, and other impairments)  may change the calculations associated with $\mathbf{P}_D$. However, this does not affect how our methodology is developed, as our framework is independent of the way in which $\mathbf{P}_D$ is calculated. We set the number of unique states experienced by each device $S=1$.}

{For the graph convolutional network (GCN) baseline, we consider an architecture consisting of an encoder with 2 convolutional layers, with 64 and 16 output channels respectively, and a 2-layer linear decoder with an output size of 16 and 1 respectively. The ReLU activation is used between each layer. The input to the GCN is a feature matrix where each row corresponds to the feature of each device, as given in Sec. VI-A. For the training process, we consider a system with $10$ devices with data from $3$ labels available at each device. To generate the training data, all possible combinations of $4$ devices are considered, and the $10$ best links are chosen based on the local reward. We then train the GCN for $100$ epochs using the Adam optimizer and a learning rate of $0.001$ and weight decay of $5 \times 10^{-4}$.}

\section{Additional Experiments}
\label{appx:exps}

\subsection{Experiments with other Datasets}
\begin{figure}[h!]	
	\centering
        \begin{subfigure}[t]{0.24\textwidth}
		\centering
            \includegraphics[width=0.99\linewidth,height=3.5cm]{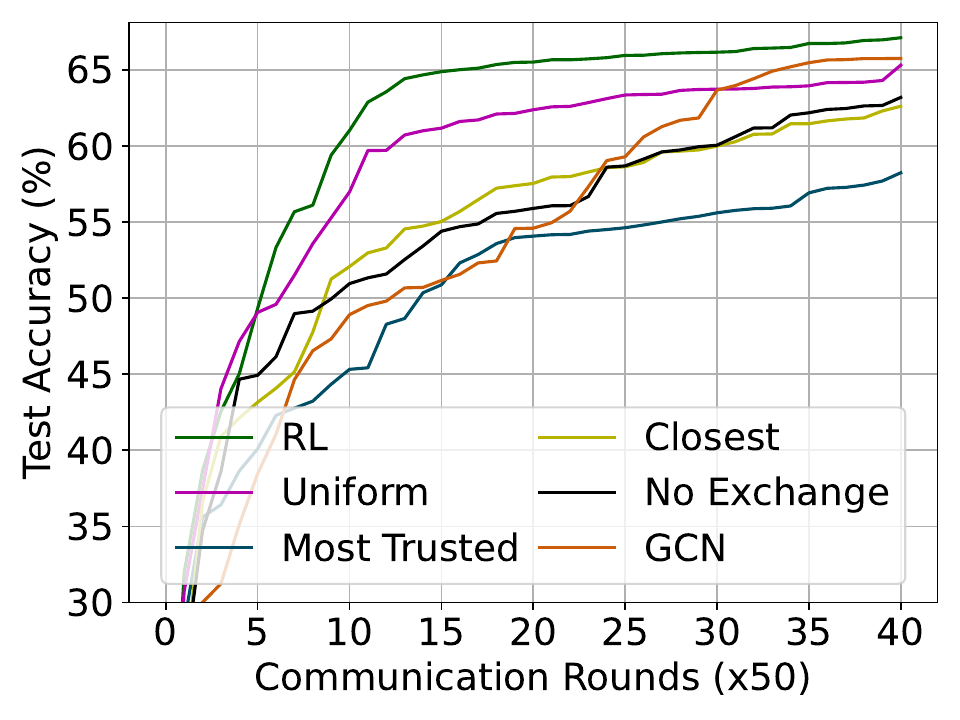}
            \vspace{-7mm}
		\caption{}\label{fig:11b}		
	\end{subfigure}
        \begin{subfigure}[t]{0.24\textwidth}
		\centering
            \includegraphics[width=0.99\linewidth,height=3.5cm]{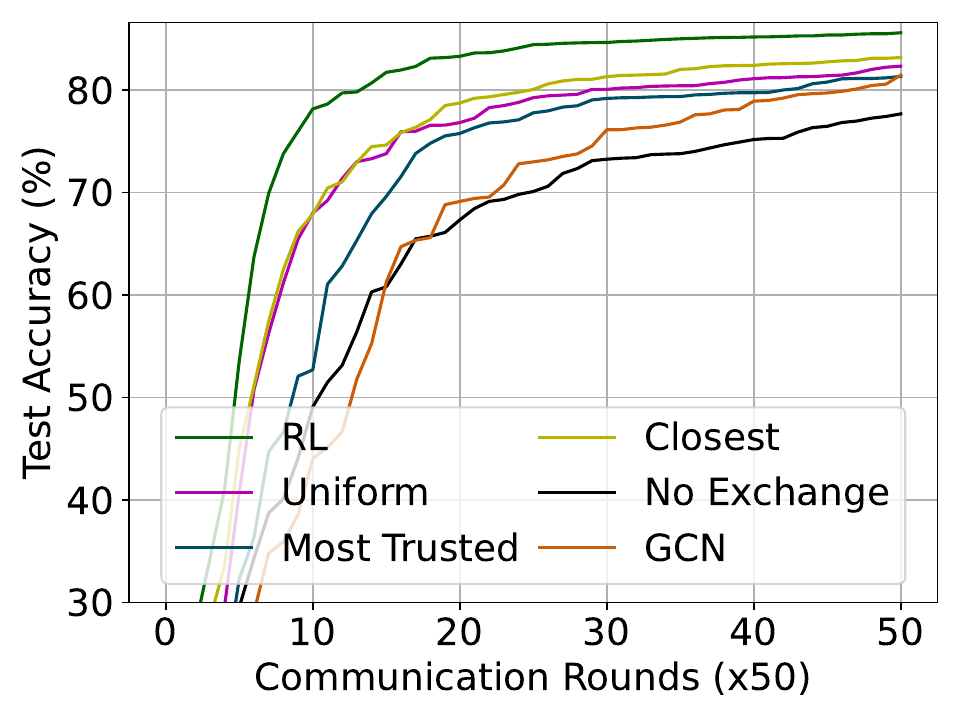} 
            \vspace{-7mm}
		\caption{}\label{fig:11c}		
	\end{subfigure}

        \begin{subfigure}[t]{0.24\textwidth}
		\centering
		\includegraphics[width=0.99\linewidth,height=3.5cm]{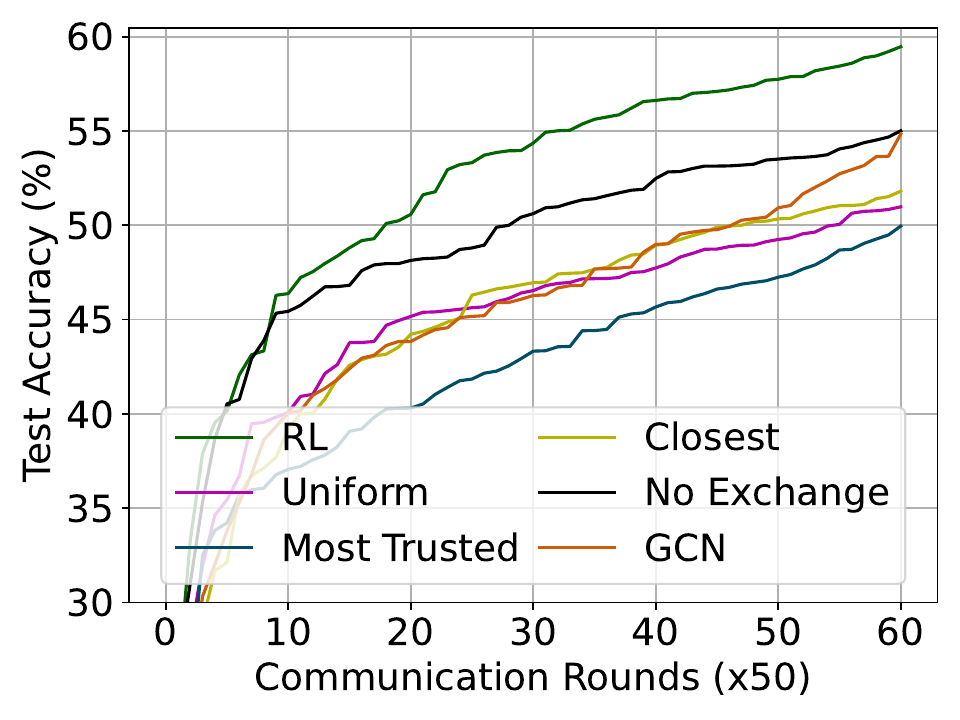}
            \vspace{-7mm}
		\caption{}\label{fig:11e}		
	\end{subfigure}
        \begin{subfigure}[t]{0.24\textwidth}
		\centering
            \includegraphics[width=0.99\linewidth,height=3.5cm]{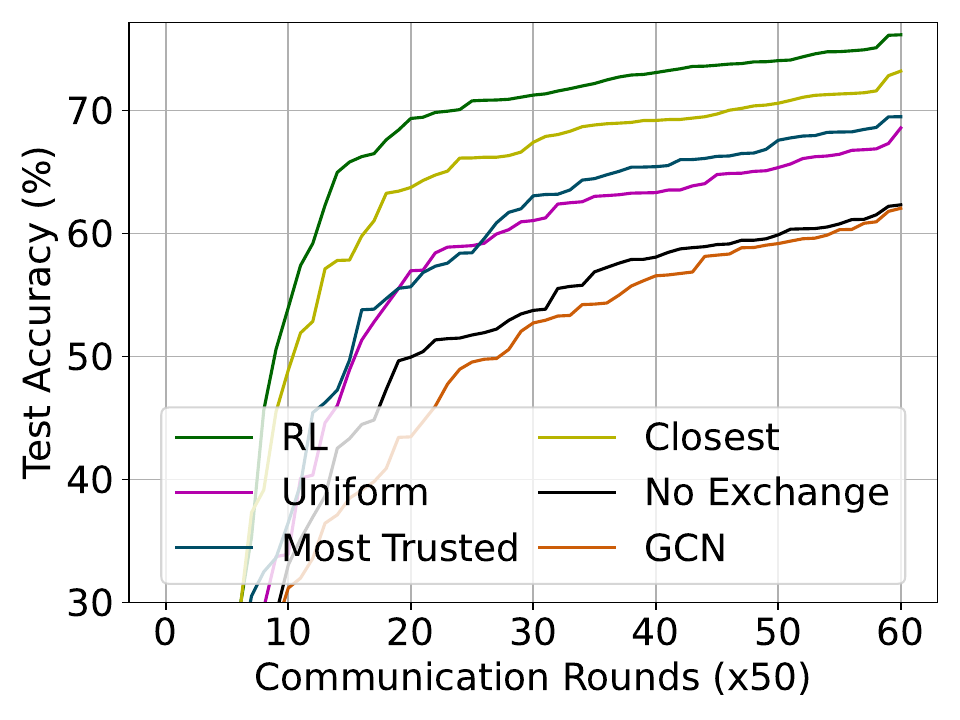} 
            \vspace{-7mm}
		\caption{}\label{fig:11g}		
	\end{subfigure}
        \begin{subfigure}[t]{0.24\textwidth}
		\centering
            \includegraphics[width=0.99\linewidth,height=3.5cm]{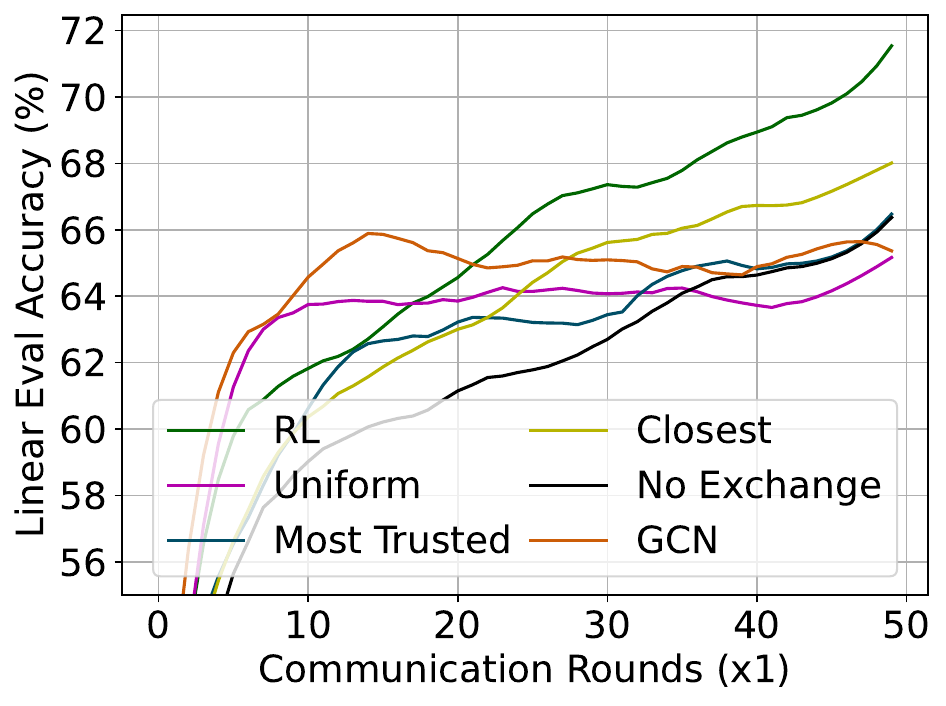} 
            \vspace{-7mm}
		\caption{}\label{fig:11f}		
	\end{subfigure}

	\caption{Our method for cooperatively discovering the optimal D2D communication graphs significantly improves performance over baselines for RadioML and SVHN datasets in the supervised (Figs. a and b), CIFAR-10 and SVHN datasets semi-supervised settings (Figs. c and d) and the USPS dataset for unsupervised settings (Fig. e). 
 }
 \label{fig:2S}\vspace{-4.7mm}
\end{figure}

Fig.~\ref{fig:2S} presents additional experiments for the setup described in Fig.~\ref{fig:2M} for the supervised setting (RadioML in (a) and SVHN in (b)) and semi-supervised setting (CIFAR-10 in (c) and SVHN in (d)). We can see that the results are qualitatively consistent with those presented in the numerical results in Sec. \ref{sec:numexp_conv}.

\begin{figure}[h!]
    \centering
    \begin{subfigure}[t]{0.24\textwidth}
        \centering
        \includegraphics[width=0.99\linewidth,height=3.5cm]{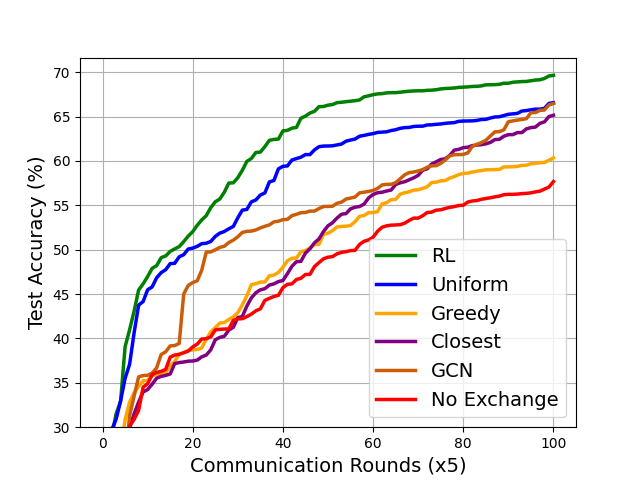}
        \vspace{-7mm}
        \caption{}\label{fig:jsd_rml}		
    \end{subfigure}
    \begin{subfigure}[t]{0.24\textwidth}
        \centering 
        \includegraphics[width=0.99\linewidth,height=3.5cm]{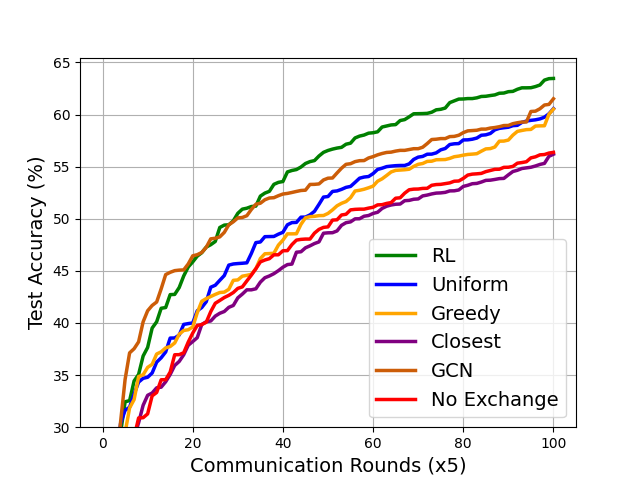}
        \vspace{-7mm}
        \caption{}\label{fig:jsd_cifar}		
    \end{subfigure}
    \begin{subfigure}[t]{0.24\textwidth}
        \centering
        \includegraphics[width=0.99\linewidth,height=3.5cm]{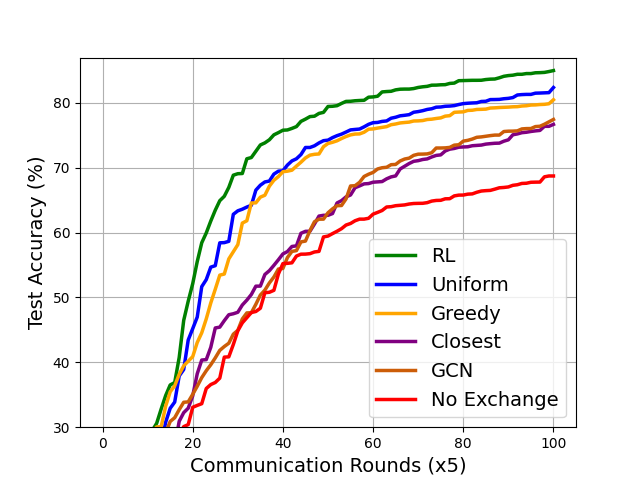}
        \vspace{-7mm}
        \caption{}\label{fig:jsd_svhn}		
    \end{subfigure}
    \caption{Our method retains performance advantages when using the Jensen-Shannon Distance instead of the 1-Wasserstein distance to calculate data diversity.}
    \label{fig:a_jsd}
\end{figure}

{\subsection{Alternative Distance Metric} 
{In Fig. \ref{fig:a_jsd}, we evaluate our method with regard to an alternative distance metric $f$ described in Sec. IV-A of the manuscript. Here, for the same supervised setting as in Fig. \ref{fig:2a}, we see that our method retains its performance advantages when using the Jensen-Shannon distance in place of the Wasserstein distance for the RadioML dataset (a), CIFAR-10 dataset (b) and SVHN dataset (c). When information is exchanged between devices to promote data diversity, the final distributions of data at each device change significantly from the original distribution. This change is captured by the Jensen-Shannon distance as well as the 1-Wasserstein distance, which increases as the initial and final distributions become more dissimilar.}}   

\begin{figure}[h!]
    \centering
    \begin{subfigure}[t]{0.24\textwidth}
        \centering
        \includegraphics[width=0.99\linewidth,height=3.5cm]{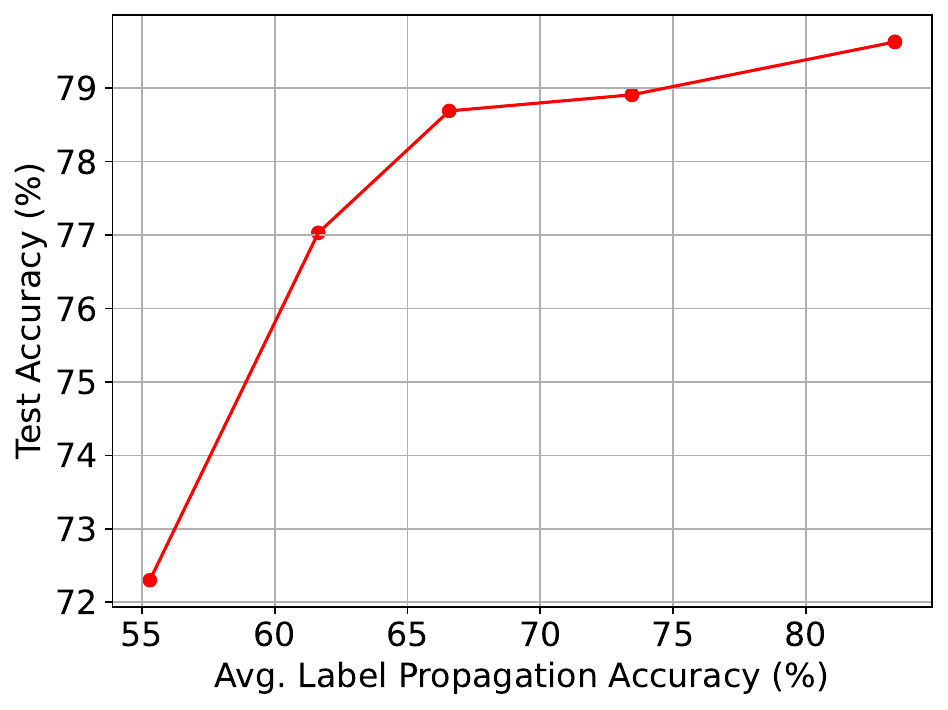}
        \vspace{-7mm}
        \caption{}\label{fig:a_lp_lc}		
    \end{subfigure}
    \begin{subfigure}[t]{0.24\textwidth}
        \centering
        \includegraphics[width=0.99\linewidth,height=3.5cm]{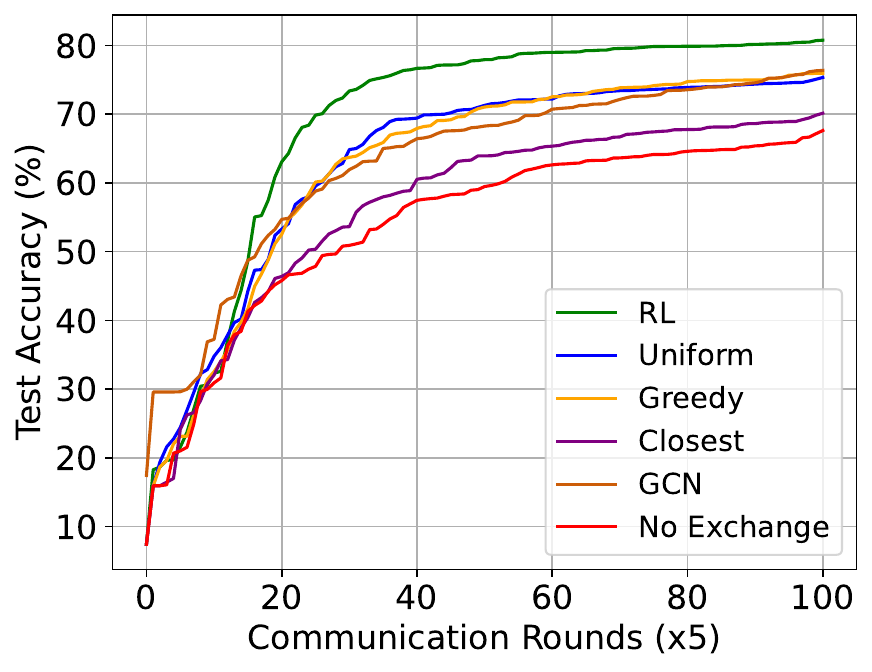}
        \vspace{-7mm}
        \caption{}\label{fig:a_lp_ssp}		
    \end{subfigure}
    \caption{The accuracy of the label propagation algorithm significantly affects the downstream federated learning performance (a), however, our method can adapt to any labeling algorithm used on unlabeled data (b).} \label{fig:a_lp}
\end{figure}

{\subsection{Impact of Label Propagation Accuracy for Semi-Supervised Scenarios} In Fig.~\ref{fig:a_lp_lc}, we analyze the impact of the label propagation accuracy on the downstream federated learning performance. We train a semi-supervised learning model on the SVHN dataset (setup from Fig.~\ref{fig:11c}), and plot the final test accuracy against the average label propagation accuracy across all devices. We vary the number of neighbors required by the kernel function of the label propagation algorithm to build a graph representation of the data, leading to differences in performance of the label propagation algorithm. We observe that as the label propagation accuracy decreases, the final test performance of federated learning decreases significantly. This is expected since a larger degree of mislabeled datapoints causes the global model to learn similar features for different labels, resulting in inaccurate classification of the testing dataset.

\subsection{Alternative Label Assignment Method for Semi-Supervised Scenarios} Next, in Fig.~\ref{fig:a_lp_ssp},we assess the performance of our method when an alternative method for labeling is used (setup from Fig.~\ref{fig:11g}. We consider a small, completely local classification model at each device to assign classes to unlabeled local data. In this method, a single-layer multi-layer perceptron (MLP) is trained on the labeled data at each device, and then used to assign classes to unlabeled data. These classes are then used as the ground truth labels. We observe that even when using the MLP to assign classes to unlabeled data, our method shows an improvement over baselines by $\sim 5\%$. This illustrates that our method is compatible with other methods for pseudo label assignment in semi-supervised learning paradigms.} 

\begin{figure}
    \includegraphics[height=5cm]{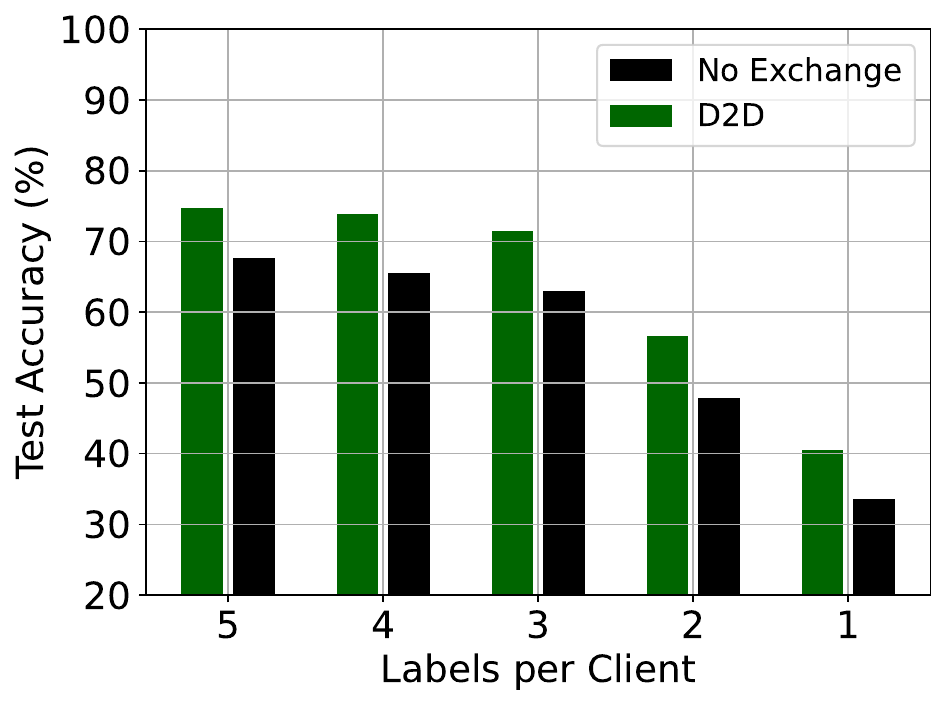}
\caption{For different levels of label skew, the relative performance improvements gained by our methods increase as the label skew between devices increases.}
\label{fig:a_skew_cifar10}
\end{figure}

\begin{figure}
    \centering
    \includegraphics[height=4.5cm]{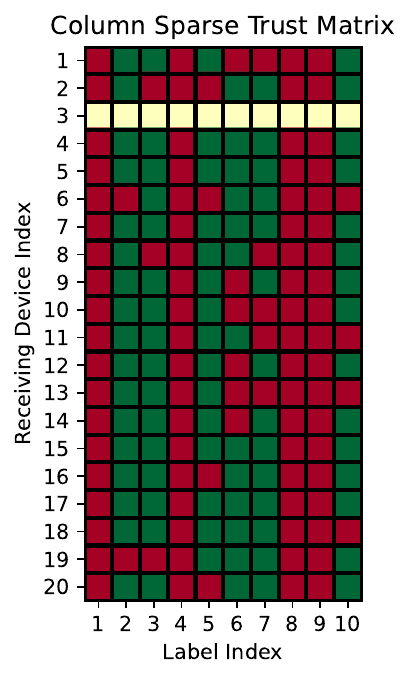}
    \includegraphics[height=4.5cm]{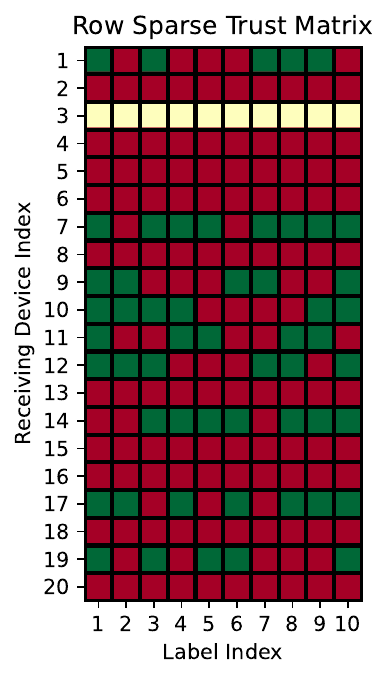}
    \includegraphics[height=4.5cm]{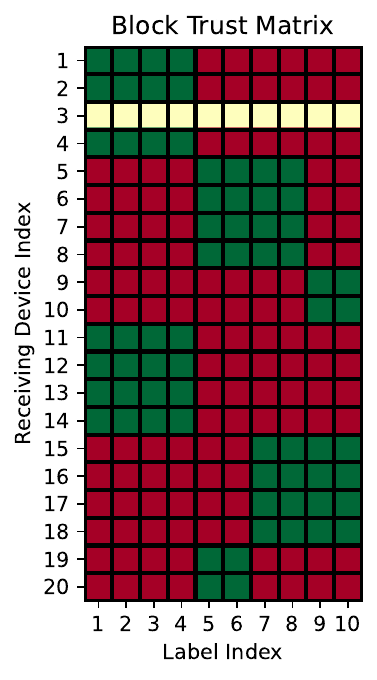}
    \caption{Example of different potential trust matrices for transmitting device $c_3$. Green entries illustrate the device-label combinations for which $c_3$ can exchange data, while red entries indicate the ones for which it cannot. Yellow squares indicate the data contained at $c_3$ itself.}\label{fig:r13_trust_matrices}
\end{figure}

\begin{figure}
    \includegraphics[height=3.25cm]{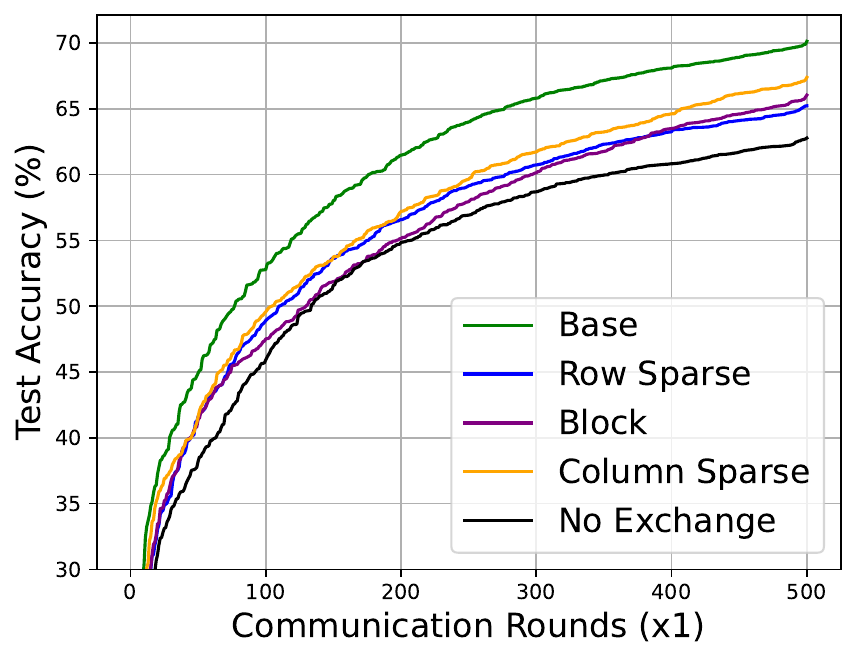}
    \includegraphics[height=3.25cm]{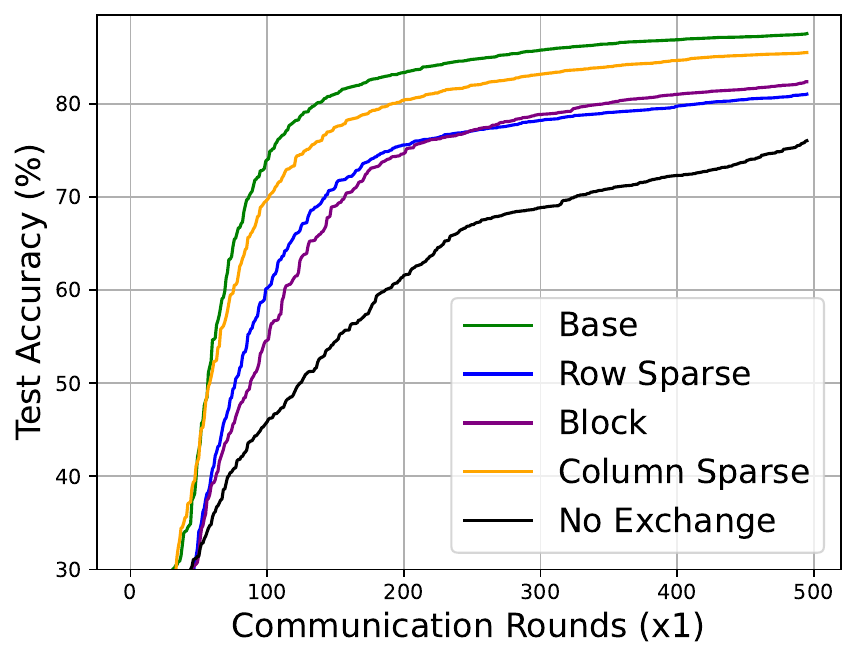}
    \caption{{The sparsity structure of the trust matrix directly impacts the extent to which the system is able to diversify local data, which impacts downstream FL performance for CIFAR-10 (left) and SVHN (right) datasets.}}\label{fig:r13_perf}
\end{figure}

{\subsection{Impact of Label Skew for Supervised Scenarios}
\label{app:skew}
In Fig.~\ref{fig:a_skew_cifar10} we observe that, for the CIFAR-10 dataset, the performance improvements of our method increase significantly with a reduction in the number of labels available at each device (increase in device skew), as seen in the $\sim 1.3 \times$ gain in test accuracy when each device has data from $1$ label before D2D exchange, compared to the $\sim 1.1 \times$ gain when each device has data from $5$ labels before D2D exchange.}

{\subsection{Impact of Structure in Trust Matrices} In Fig.~\ref{fig:r13_trust_matrices}, we give examples of different sparsity structures that may be present in trust matrices. In Fig.~\ref{fig:r13_perf}, we assess the impact of these different sparsity structures on downstream FL performance, conducting experiments using block, row sparse, and column sparse trust matrices with the CIFAR-10 (left) and SVHN datasets (right) using $20$ devices. We set all elements in randomly selected rows or columns to $0$ to create row sparse and column sparse matrices respectively. Using this process, we create trust matrices with $50\%$ row sparsity and $50 \%$ column sparsity. For the block matrix case, we populate the trust matrix of each device with blocks of sizes between $2$ and $4$ chosen at random. We compare the effects of these trust matrix structures on our method to a baseline trust matrix that has been randomly generated. We observe that in both cases, there is a noticeable performance gap between a randomly generated trust matrix and block, row sparse or column sparse trust matrices. Intuitively, row sparsity forbids the transmitting device from sharing any information with the corresponding devices, restricting the options for available for a selected requesting device. In a column sparse matrix, a transmitting device cannot share information from the corresponding partitions with any requesting device. This reduces the data diversity improvements achieved by requesting devices after requesting information from the transmitting device. In a block matrix, each device can only share a limited amount of information with requesting devices, and is also restricted in terms of the choice of devices to request information from. This has a significant effect on the achievable data diversity for both requesting and transmitting data. We also observe that the column sparse trust matrix has better performance than the row sparse and block matrices. The impact of column sparsity is less pronounced for learning tasks with highly skewed partitions, such as the ones used in our experiments, as columns corresponding to partitions for which the transmitter has no data have no effect on the performance. }

\begin{figure}
    \centering
    \includegraphics[height=5cm]{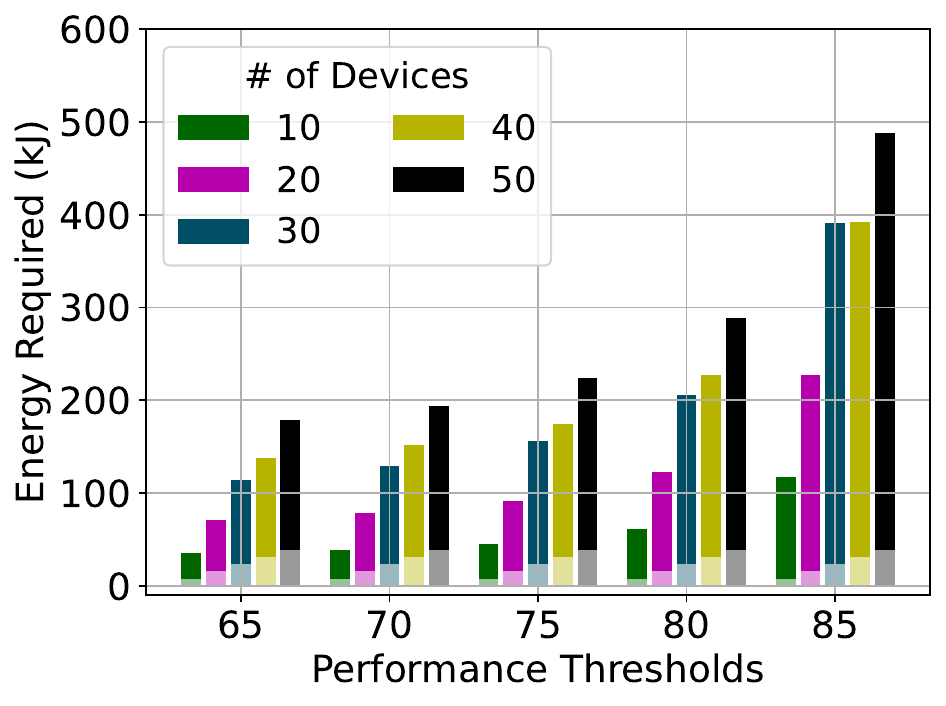}
    \caption{Energy consumption for varying system sizes and performance thresholds on the supervised setting with SVHN dataset. The overall communication overhead of our method scales linearly with the number of devices for a given performance threshold to be reached.}
    \label{fig:energy_svhn}
\end{figure}

{\subsection{Energy Consumption for Varying System Sizes} In Fig.~\ref{fig:energy_svhn}, we observe that our method generalizes well to the SVHN dataset as well, with the total communication overhead scaling linearly with the number of devices for a given performance threshold to be reached, thus retaining the advantages of our method as the system size increases.}

\end{appendices}

\end{document}